%% file: multib_paper.tex
\pdfoutput=1
\newcommand*{\ATLASLATEXPATH}{}
\documentclass[PAPER,cernpreprint,UKenglish,texlive=2011,txfonts, texmf]{\ATLASLATEXPATH atlasdoc}
\usepackage[subfigure=true]{\ATLASLATEXPATH atlaspackage}
\usepackage{\ATLASLATEXPATH atlasbiblatex}
\usepackage{\ATLASLATEXPATH atlascontribute}
\usepackage[BSM=true]{\ATLASLATEXPATH atlasphysics}

\graphicspath{{}}

\usepackage{multib_paper-defs}
\usepackage{arydshln} 
\usepackage{multirow} 
\usepackage{pdflscape}

\AtlasTitle{Search for pair production of gluinos decaying via stop
and sbottom
in events with $b$-jets and large missing transverse
momentum in $pp$ collisions at $\sqrt{s}=13 \tev$ with the ATLAS detector}
\author{The ATLAS Collaboration}

\hypersetup{pdftitle={ATLAS draft},pdfauthor={The ATLAS Collaboration}}

\PreprintIdNumber{CERN-EP-2016-061}

 \AtlasJournal{Phys.\ Rev.\ D.}

\AtlasAbstract{%
A search for Supersymmetry involving the pair production of gluinos decaying via 
third-generation squarks to the lightest neutralino ($\displaystyle\tilde\chi^0_1$) is
reported. It uses an LHC  proton--proton
dataset at a center-of-mass energy $\sqrt{s} = 13 \tev$  with
an integrated luminosity of 3.2~\ifb\ collected with the ATLAS
detector in 2015. The signal is searched for in
events containing several energetic jets, of which at least three must
be identified as $b$-jets, large missing transverse momentum and,
potentially, isolated electrons or muons. Large-radius jets with a
high mass are also used to identify highly boosted top quarks. 
No excess is found above the predicted background.  
For $\displaystyle\tilde\chi^0_1$ masses below approximately 700~\gev, gluino masses of less than 1.78~\tev\ and 1.76~\tev\ are excluded at
the 95\%~CL in simplified models of the pair production of gluinos decaying via sbottom and stop,
respectively. These results significantly extend the exclusion limits obtained with the $\rts = 8 \tev$ dataset. 

}

\begin{document}

\maketitle

\tableofcontents

\section{Introduction}
\label{sec:intro}

Supersymmetry (SUSY)
\cite{Golfand:1971iw,Volkov:1973ix,Wess:1974tw,Wess:1974jb,Ferrara:1974pu,Salam:1974ig}
is a generalization of space-time symmetries that 
predicts new bosonic partners to the fermions and new fermionic partners to the bosons
of the Standard Model (SM). 
If $R$-parity is conserved~\cite{Farrar:1978xj},
SUSY particles are produced in pairs and the lightest supersymmetric particle (LSP) is stable. 
The scalar partners of  the left- and right-handed quarks, the squarks $\squarkL$ and $\squarkR$,
can mix to form two mass eigenstates $\tilde{q}_1$ and $\tilde{q}_2$, ordered by increasing mass. 
SUSY can solve the hierarchy problem \cite{Sakai:1981gr,Dimopoulos:1981yj,Ibanez:1981yh,Dimopoulos:1981zb} by preventing “unnatural” fine-tuning in the
Higgs sector provided that the superpartners of the top quark (stop, $\stopone$ and $\stoptwo$) have masses not too far
above the weak scale. 
Because of the  SM weak isospin symmetry, the mass of the left-handed bottom quark scalar partner (sbottom, $\sbottomL$) is tied 
to the mass of the left-handed top quark scalar partner ($\stopL$), and as a consequence the mass of the lightest sbottom $\sbottomone$ is also 
expected to be close to the weak scale. The fermionic partners of the gluons, 
the gluinos ($\gluino$), are also constrained by naturalness \cite{Barbieri:1987fn,deCarlos:1993yy} to have a mass around the \tev\ scale in
order to limit their contributions to the radiative corrections to the stop masses. For these reasons, and because the gluinos are expected to be pair-produced with a high cross-section at the Large Hadron Collider (LHC),
the search for gluino production with decays via stop and sbottom quarks is highly motivated at the LHC.

\begin{figure*}[h]
\centering 
\subfigure[]{\includegraphics[width=0.35\textwidth]{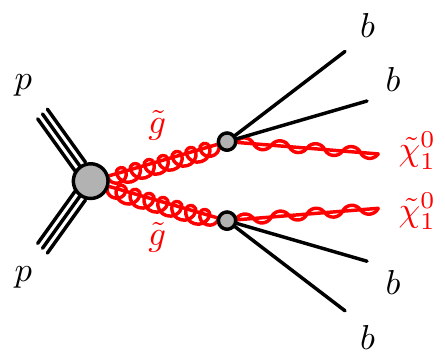}}
\subfigure[]{\includegraphics[width=0.35\textwidth]{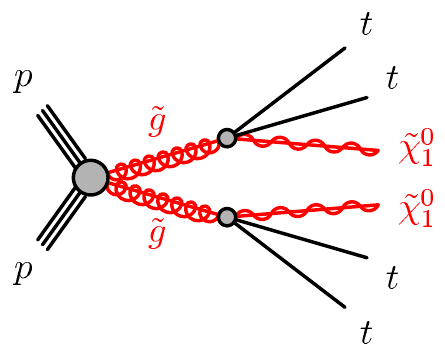}}\hspace{0.05\textwidth}
\caption{The decay topologies in the (a) Gbb and (b) Gtt simplified models.
}\label{fig:feynman-Gluino-topbottom-offshell}
\end{figure*}

This paper presents the search for gluino pair production where both gluinos either decay to stops via $\gluino \to \stopone t$, or to sbottoms via $\gluino \to \sbottomone b$, using a dataset of 3.2 fb$^{-1}$ of proton--proton data collected with the ATLAS detector~\cite{Aad:2008zzm} at a center-of-mass energy of $\sqrt{s} = 13 \tev$. 
Each stop (sbottom) is then assumed to decay to a top (bottom) quark and the LSP: $\stopone \to t \ninoone$ ($\sbottomone \to b \ninoone$). The LSP is assumed to 
be the lightest neutralino $\ninoone$, the lightest linear superposition of the
superpartners of the neutral electroweak and Higgs bosons. The $\ninoone$  interacts only weakly, resulting in final states with substantial missing transverse momentum of magnitude $\met$.
Diagrams of the simplified models \cite{Alwall:2008ag,Alves:2011wf} considered, which are referred to as ``Gbb'' and ``Gtt'' in the following, are shown in Figures~\ref{fig:feynman-Gluino-topbottom-offshell}(a) and \ref{fig:feynman-Gluino-topbottom-offshell}(b), respectively.
The sbottom and stop are assumed to be produced off-shell such that the gluinos undergo the three-body decay $\gluino \to b\bar{b}\ninoone$ or $\gluino \to t\bar{t}\ninoone$, 
and that the only parameters of the simplified models are the gluino and $\ninoone$ masses.\footnote{Models with on-shell sbottom and stop were 
studied in Run 1 \cite{3bjetsPaper} and the limits on the gluino and the $\ninoone$ masses were found to be mostly independent of the stop and sbottom masses, except when the stop is very light.}

The Gbb experimental signature consists of four energetic $b$-jets (i.e. jets containing $b$-hadrons) and large $\met$. In order to maintain high signal efficiency, at least three of four required jets must be identified as $b$-jets ($b$-tagged). This requirement is very effective in rejecting $\ttbar$ events, which constitute the main background for both the Gbb and Gtt signatures, and which contain only two $b$-jets unless they are produced with additional heavy-flavor jets. The Gtt experimental signature also contains four $b$-jets and $\met$, but yields in addition four $W$ bosons originating 
from the top quark decays $t \to W b$. Each $W$ boson can either decay leptonically ($W \to \ell \nu$) or hadronically ($W \to q \bar{q}^\prime$).
A Gtt event would therefore possess a high jet multiplicity, with as many as 12 jets originating from top quark decays and, potentially, isolated charged leptons. In this paper, pair-produced gluinos decaying via stop and sbottom quarks are searched for using events with high jet multiplicity, of which at least three must be identified as $b$-jets, large $\met$, and either zero leptons (referred to as Gtt 0-lepton channel) or at least one identified charged lepton\footnote{The term ``lepton'' refers exclusively to an electron or a muon in this paper.} (referred to as Gtt 1-lepton channel). 
For both the Gbb and Gtt models, several signal regions are designed to cover different ranges of gluino and $\ninoone$ masses. For the Gtt models with a large mass difference (mass splitting) between the gluino and $\ninoone$, the 
top quarks tend to be highly boosted and their decay products collimated. 
In the corresponding signal regions, at least one large-radius, trimmed \cite{Krohn:2009th} jet, which is re-clustered from small-radius jets \cite{Nachman:2014kla}, is required to have a high mass to identify hadronically decaying boosted top quarks.  

Pair production of gluinos, with subsequent decays via sbottom quarks, was searched for in ATLAS Run 1 with a similar analysis requiring at least three $b$-tagged jets \cite{3bjetsPaper}. It excluded gluino masses below 1290 \gev\ for LSP masses below 400~\gev\ at 95\% confidence level (CL). That analysis also searched for gluinos decaying via stop quarks in events with at least three $b$-tagged jets and either zero or at least one identified lepton and obtained the best ATLAS
limits for the Gtt models with massless and moderately massive LSP \cite{Aad:2015iea}. Gluino masses below 1400 \gev\ were excluded at 95\%~CL for LSP masses below 400 \gev.
Pair-produced gluinos with stop-mediated decays have also been searched for by ATLAS in events with high jet multiplicity \cite{multijetsPaper}, events with at least one lepton, many jets, and $\met$ \cite{1lepPaper}, and 
 events containing pairs of 
same-sign leptons or three leptons \cite{SS3LPaper}, the latter obtaining the best ATLAS limit for Gtt models with compressed mass spectra between the gluino and the LSP \cite{Aad:2015iea}. 

Similar searches performed in the CMS experiment \cite{Chatrchyan:2013wxa, Chatrchyan:2013iqa, Khachatryan:2015pwa, Chatrchyan:2014lfa, Chatrchyan:2013fea, CMS:2014dpa} have produced comparable results to ATLAS searches.

\section{ATLAS detector}
\label{sec:detector}
The ATLAS detector is a multipurpose particle physics detector with a forward-backward 
symmetric cylindrical geometry and nearly 4$\pi$ coverage in solid angle.\footnote{ATLAS 
uses a right-handed coordinate system with its origin at the nominal interaction point in 
the center of the detector. The positive $x$-axis is defined by the direction from the 
interaction point to the center of the LHC ring, with the positive $y$-axis pointing 
upwards, while the beam direction defines the $z$-axis. Cylindrical coordinates $(r,\phi)$ 
are used in the transverse plane, $\phi$ being the azimuthal angle around the $z$-axis. 
The pseudorapidity $\eta$ is defined in terms of the polar angle $\theta$ by $\eta=-\ln\tan(\theta/2)$.}
The inner tracking detector (ID) consists of pixel and silicon microstrip detectors 
covering the pseudorapidity region $|\eta|<2.5$, surrounded by a transition radiation tracker, 
which enhances electron identification in the region $|\eta|<2.0$. 
Before the start of Run 2, the new innermost pixel layer, the Insertable B-Layer (IBL) \cite{Capeans:1291633}, was inserted at a mean sensor radius of 3.3~cm.
The ID is surrounded by a thin superconducting solenoid providing an axial 2 T magnetic field and by
a fine-granularity lead/liquid-argon (LAr) electromagnetic calorimeter covering $|\eta|<3.2$.
A steel/scintillator-tile calorimeter provides coverage for hadronic showers in
the central pseudorapidity range ($|\eta|<1.7$). 
The endcap and forward regions ($1.5<|\eta|<4.9$) of the hadronic calorimeter are made of LAr active layers with either copper or tungsten as the absorber material. 
A muon spectrometer with an air-core toroid magnet system surrounds the calorimeters.
Three layers of high-precision tracking chambers
provide coverage in the range $|\eta|<2.7$, while dedicated fast chambers allow triggering in the region $|\eta|<2.4$.
The ATLAS trigger system \cite{atlastrigger} consists of a hardware-based Level-1 trigger followed by a software-based High Level Trigger.

\section{Data and simulated event samples}
\label{sec:samples}
The data used in this analysis were collected by the ATLAS detector from $pp$ collisions produced by the 
LHC at a center-of-mass-energy of 13 \tev\ and 25 ns proton bunch spacing. The full dataset 
corresponds to an integrated luminosity of 3.2 \ifb\, with an associated uncertainty of $\pm$5\%, after 
requiring that all detector subsystems were operational during data recording. The measurement of 
the integrated luminosity is derived, following a methodology similar to that detailed in Ref.~\cite{Aad:2013ucp}, 
from a calibration of the luminosity scale using a pair of $x$--$y$ beam-separation scans performed in June 2015.
Events are required to pass an $\met$ trigger that is fully efficient for events passing the preselection defined in Section~\ref{sec:evsel}. 
Each event includes on average 14 additional inelastic $pp$ collisions (``pileup'') in the same bunch crossing. 

Simulated event samples are used to model the signal and background processes in this analysis. The signal samples 
for the Gbb and Gtt processes are generated with up to two additional partons using \MGMCatNLO~\cite{amcatnlo_mg5} v2.2.2 
at leading order (LO) with CTEQ6L1~\cite{Nadolsky:2008zw} Parton Density Function (PDF) sets and interfaced to \PYTHIA v8.186~\cite{pythia8} 
for the modeling of the parton showering, hadronization and underlying event. 

The dominant background in the signal regions is the production of \ttbar pairs with additional high-\pt 
jets. The sample for the estimation of this background is generated using the \POWHEGBOX~\cite{Alioli:2010xd, Frixione:2007vw} 
generator at next-to-leading order (NLO) with CT10~\cite{Lai:2010vv} PDFs and interfaced to \PYTHIA v6.428~\cite{pythia} for 
showering and hadronization. 
The decays of heavy-flavor hadrons are modeled using the \EVTGEN~\cite{Lange:2001uf} package.
The $h_\mathrm{damp}$ parameter in \POWHEG, which controls the \pt of the first additional 
emission beyond the Born level and thus regulates the \pt of the recoil emission against the \ttbar system, 
is set to the mass of the top quark ($m_\mathrm{top} = $ 172.5 \gev). This setting was found to give the best 
description of the \pt of the \ttbar system at $\sqs=7
\tev$~\cite{hdamp-proof} and $\sqs
=8 \tev$~\cite{Aad:2015hna}. 
All events with at least one semileptonically decaying top quark are included. Fully hadronic \ttbar events 
do not contain sufficient \met\ to contribute significantly to the background. 
    
Smaller backgrounds in the signal region come from the production of
\ttbar pairs in association with $W/Z/h$ and additional jets, 
single-top production, production of $t\bar{t}t\bar{t}$, $W/Z+$jets and $WW$/$WZ$/$ZZ$ (diboson) events. The 
production of \ttbar pairs in association with electroweak vector bosons and $\ttbar \ttbar$ production are modeled by 
samples generated using \MADGRAPH~\cite{madgraph} interfaced to \PYTHIA v8.186, while samples to model ${\ttbar}h$ production 
are generated using \MGMCatNLO~\cite{amcatnlo_mg5} v2.2.1 and showered with \MYHERWIG++~\cite{Bahr:2008pv} 
v2.7.1. Single-top production in the $s$-, $t$- and $Wt$-channel are generated by \POWHEGBOX interfaced to 
\PYTHIA v6.428. $W/Z$+jets and diboson processes are simulated using the \SHERPA v2.1.1~\cite{Gleisberg:2008ta} 
generator with CT10 PDF sets. Matrix elements for these processes are calculated using the Comix~\cite{Gleisberg:2008fv} 
and OpenLoops~\cite{Cascioli:2011va} generators and merged with the \SHERPA parton shower~\cite{Schumann:2007mg} 
using the ME$+$PS$@$NLO prescription~\cite{Hoeche:2012yf}. 

All simulated event samples, with the exception of the Gbb signals, are passed through full 
ATLAS detector simulation using \MYGEANT4~\cite{geant4,atlassimulation}. The Gbb signal samples are 
passed through a fast simulation that uses a parameterized description to simulate the response of the 
calorimeter systems~\cite{atlfast}. The simulated events are reconstructed with the same algorithm as 
that used for data. All \PYTHIA v6.428 samples use the PERUGIA2012~\cite{perugia} set of tuned parameters (tune) 
for the underlying event, while \PYTHIA v8.186 and \MYHERWIG++ showering are run with the A14~\cite{a14} 
and UEEE5~\cite{ueee5} underlying-event tunes, respectively. In-time and out-of-time pileup interactions 
from the same or nearby bunch-crossings are simulated by overlaying additional $pp$ collisions generated by 
\PYTHIA v8.186 on the hard-scattering events. Details of the sample generation and normalization are 
summarized in Table~\ref{t-MCsamples}. Additional samples with different generators and settings are used 
to estimate systematic uncertainties on the backgrounds, as described in Section~\ref{sec:syst}. 

The signal samples are normalized using the best cross-sections calculated at NLO in the strong coupling 
constant, adding the resummation of soft gluon emission at next-to-leading-logarithmic (NLL) accuracy
~\cite{Beenakker:1996ch,Kulesza:2008jb,Kulesza:2009kq,Beenakker:2009ha,Beenakker:2011fu}. The nominal cross-section 
and the uncertainty are taken from an envelope of cross-section predictions using different PDF sets and factorization 
and renormalization scales, as described in Ref.~\cite{Kramer:2012bx}. The cross-section of gluino pair-production 
in these simplified models is approximately 325 fb for a gluino mass of 1~\tev, falling to 2.8 fb for 1.8~\tev\
mass gluinos. All background processes are normalized to the best available theoretical calculation for their 
respective cross-sections. The order of this calculation in perturbative QCD (pQCD) for each process is listed in 
Table~\ref{t-MCsamples}.

\begin{table}[h]

\centering
{\footnotesize
\begin{tabular}{ccccc}
\toprule
Process &    Generator                   & Tune & PDF set  &  Cross-section  \\
        & + fragmentation/hadronization  &      &          &  order \\
\midrule
{\bf \ttbar} & \POWHEGBOX v2
& {\textsc PERUGIA2012} & {\textsc CT10} & NNLO+NNLL~\cite{Czakon:2011xx} \\
  & + \PYTHIA-6.428 & & & \\
\\
{\bf Single top} & \POWHEGBOX v2
& {\textsc PERUGIA2012} & {\textsc CT10} & NNLO+NNLL~\cite{Kidonakis:2011wy,Kidonakis:2010ux,Kidonakis:2010tc}  \\
  & + \PYTHIA-6.428 & & & \\
\\
{\bf ${\ttbar}W$/${\ttbar}Z$/4-tops} & \MADGRAPH-2.2.2
& {\textsc A14} & {\textsc NNPDF2.3~\cite{Ball:2012cx}} & NLO \\
  & + \PYTHIA-8.186 & & & \\
\\
{\bf ${\ttbar}h$} & \MGMCatNLO-2.2.1 
& {\textsc UEEE5} & {\textsc CT10} & NLO~\cite{Heinemeyer:1559921} \\
  & + \MYHERWIG++-2.7.1 & & & \\

\\
{\bf Dibosons}  & \SHERPA-2.1.1 & Default & {\textsc CT10} & NLO \\
      $WW$, $WZ$, $ZZ$ & & & & \\
\\
{\bf $W$/${Z}+$jets} & \SHERPA-2.1.1 & Default & {\textsc CT10} & NNLO~\cite{Catani:2009sm}  \\
\bottomrule
\end{tabular}
}
\caption{List of generators used for the different background
processes. Information is given about the pQCD highest-order accuracy
used for the normalization of the different samples, the underlying-event 
tunes and PDF sets considered.}
   \label{t-MCsamples}
\end{table}

\section{Object reconstruction}
\label{sec:objdef}
Interaction vertices from the proton--proton collisions are reconstructed from at least two tracks with 
$\pt > 0.4 \gev$, and are required to be consistent with the beamspot envelope. The primary vertex is 
identified as the one with the largest sum of squares of the transverse momenta from associated 
tracks ($\sum{|p_{\mathrm{T,track}}|^{2}}$)~\cite{ref-pv}.

Basic selection criteria are applied to define candidates for electrons, muons and jets in the event. An 
overlap removal procedure is applied to these candidates to prevent double-counting. 
Further requirements are then made to select the final signal leptons and jets 
from the remaining objects. The details of the object selections and of the overlap removal procedure 
are given below. 

Candidate jets are reconstructed from three-dimensional topological energy clusters~\cite{topoclusters} in 
the calorimeter using the anti-$k_{t}$ jet algorithm~\cite{Cacciari:2008gp} with a radius parameter of 0.4 (small-$R$ jets). 
Each topological cluster is calibrated to the electromagnetic scale response prior to jet reconstruction. The 
reconstructed jets are then calibrated to the particle level by the application of a jet energy scale 
(JES) derived from simulation and corrections based on 8~\tev\ data~\cite{Run2-JES,Final-Run1-JES}. 
Quality criteria are imposed to reject events that contain at least one jet arising from non-collision sources 
or detector noise~\cite{Run2-jet-cleaning}. Further selections are applied 
to reject jets that originate from pileup interactions~\cite{Aad:2015ina}. Candidate jets are 
required to have \pt$>$~20 \gev\ and $|\eta| < 2.8$. Signal jets, selected after resolving overlaps with electrons 
and muons, are required to satisfy the stricter requirement of \pt$>$~30~\gev. 

A multivariate algorithm using information about the impact 
parameters of inner detector tracks matched to the jet, the presence of displaced secondary vertices, and the 
reconstructed flight paths of $b$- and $c$-hadrons inside the jet~\cite{btag-Run2,btag-Run1,btag-mistag-Run1} is used to tag $b$-jets. The $b$-tagging 
working point with an 85\% efficiency, as determined from a simulated sample of \ttbar events, was found to be optimal.
The corresponding rejection factors against jets originating from $c$-quarks, from $\tau$-leptons and from light 
quarks and gluons in the same sample at this working point are 2.6, 3.8 and 27, respectively. 

The candidate small-$R$ jets are used as inputs for further jet re-clustering~\cite{Nachman:2014kla} using the 
anti-$k_{t}$ algorithm with a radius parameter of 1.0. These re-clustered jets are then trimmed~\cite{Krohn:2009th,Nachman:2014kla} 
by removing subjets whose \pt falls below $f_\mathrm{cut}=$ 5\% of the \pt\ of the original re-clustered jet. The resulting 
large-$R$ jets are used to tag high-\pt boosted top quarks in the event. Selected large-$R$ jets are required to have 
\pt$>$~300 \gev\ and to have $|\eta|<$~2.0. A large-$R$ jet is tagged as a top candidate if it has a mass above 
100~\gev. When it is not explicitly stated otherwise, the term ``jets'' in this paper refers to small-$R$ jets.

Electron candidates are reconstructed from energy clusters in the electromagnetic calorimeter and inner 
detector tracks and are required to satisfy a set of ``loose'' quality criteria~\cite{Aad:2014fxa,Aad:2014ephr,eID-Run2}. 
They are also required to have $|\eta|<$~2.47. Muon candidates are reconstructed from matching tracks in 
the inner detector and in the muon spectrometer. They are required to meet ``medium'' quality criteria, 
as described in Refs.~\cite{muID-Run2,ref-mu-Run1} and to have $|\eta|<$~2.5. All electron and muon 
candidates must have \pt$>$~20 \gev\ and survive the overlap removal procedure. Signal leptons are chosen 
from the candidates with the following isolation requirement -- the scalar sum of \pt of additional inner 
detector tracks in a cone around the lepton track is required to be $<$5\% of the lepton \pt. The angular 
separation between the lepton and the $b$-jet ensuing from a semileptonic top quark decay narrows as 
the \pt of the top quark increases. This increased collimation is accounted for by varying the radius 
of the isolation cone as max(0.2, 10/$p_{\mathrm{T}}^{\mathrm{lep}}$), where $p_{\mathrm{T}}^{\mathrm{lep}}$ 
is the lepton \pt expressed in~\gev. Signal electrons are further required to meet the ``tight'' quality 
criteria, while signal muons are required to satisfy the same ``medium'' quality criteria as the muon candidates. 
Electrons (muons) are matched to the primary vertex by requiring the transverse impact parameter $d_{0}$ to 
satisfy $|d_{0}|/\sigma(d_{0})<$~5 (3), where $\sigma(d_{0})$ is the measured uncertainty in $d_{0}$, 
and the longitudinal impact parameter $z_{0}$ to satisfy $|z_{0}\sin\theta|<$~0.5 mm. In addition, events 
containing one or more muon candidates with $|d_{0}|$ ($|z_{0}|$) $>$ 0.2 mm (1 mm) are rejected to suppress 
cosmic rays. 

The overlap removal procedure between muon and jet candidates is designed to remove those muons 
that are likely to have originated from the decay of hadrons and to retain the overlapping jet. 
Jets and muons may also appear in close proximity when the jet results from high-\pt muon bremsstrahlung, and 
in such cases the jet should be removed and the muon retained. Such jets are characterized by having very 
few matching inner detector tracks. Therefore, if the angular distance ${\Delta}R$ between a muon and a jet is 
within min(0.4, 0.04 + 10~\gev/\pt) of the axis of a jet,\footnote{${\Delta}R = \sqrt{(\Delta\eta)^{2} + (\Delta\phi)^{2}}$ defines the distance between objects 
in ($\eta$,$\phi$) space.} the muon is removed only if the jet has $\ge$3 
matching inner detector tracks. If the jet has fewer than three matching tracks, the jet is removed and the muon is kept~\cite{ref-ttres-muOR}.
Overlap removal between electron and jet candidates aims to remove jets that are formed primarily from 
the showering of a prompt electron and to remove electrons that are produced in the decay chains of hadrons. Since 
electron showers within the cone of a jet contribute to the measured energy of the jet, any overlap between an electron 
and the jet must be fully resolved. A \pt-dependent cone for the purpose of this overlap removal is thus impractical. 
Consequently, any non-$b$-tagged jet whose axis lies ${\Delta}R < 0.2$ from an electron is discarded. If 
the electron is within ${\Delta}R = 0.4$ of the axis of any jet remaining after this initial overlap removal procedure, 
the jet is retained and the electron is removed. Finally, electron candidates that lie ${\Delta}R < 0.01$ from muon 
candidates are removed to suppress contributions from muon bremsstrahlung. 

The missing transverse momentum ($\met$) in the event is defined as the magnitude of the negative vector sum transverse momentum ($\vec{\pt}^{\mathrm{miss}}$) of 
all selected and calibrated objects in the event, with an extra term added to account for soft 
energy that is not associated to any of the selected objects. This soft term is calculated from 
inner detector tracks matched to the primary vertex to make it more resilient to contamination from pileup interactions ~\cite{MET-Run2,MET-Run1} .

Corrections derived from data control samples are applied to simulated events to account for differences between 
data and simulation in the reconstruction efficiencies, momentum scale and resolution of leptons
~\cite{Aad:2014ephr,eID-Run2,muID-Run2,ref-mu3-Run1} and in the efficiency and false positive rate for identifying $b$-jets~\cite{btag-mistag-Run1,btag-Run1}.

\section{Event selection}
\label{sec:evsel}
The event selection criteria are defined based on kinematic requirements on the objects defined 
in Section~\ref{sec:objdef} and on the following event variables.

Two effective mass variables are used, which would typically have much higher
values in pair-produced gluino events than in background events. The Gtt signal regions employ the
inclusive effective mass $\meffi$:
\begin{equation*}
 \meffi = \sum_{i} {\pt}^{\mathrm{jet}_{i}}  + \sum_{j} {\pt}^{\ell_j}  + \met,
\end{equation*}
where the first and second sums are over the signal jets and leptons,
respectively. The signal regions for the Gbb models, for which four high-$\pt$
$b$-jets are expected, are defined using $\meffe$:
\begin{equation*}
\meffe = \sum_{i\leq 4} {\pt}^{\mathrm{jet}_{i}} + \met,
\end{equation*}
where the sum is over the four highest-$\pt$ (leading) signal jets in the event.

In regions with at least one signal lepton, the transverse mass
$\mt$
of the leading signal lepton ($\ell$) and $\met$ is used to
discriminate between the signal and backgrounds from semileptonic
$\ttbar$ and $W$+jets events:
\begin{equation*}
\mt = \sqrt{2\pt^\ell \met\{1-\cos[\Delta\phi(\vec{\pt}^{\mathrm{miss}},\ell)]\}}.
\end{equation*}
Neglecting resolution effects, $\mt$ is bounded from above by the $W$
boson mass for these backgrounds and typically has higher values for
Gtt events. 
 Another useful
transverse mass variable is $\mtb$, the minimum transverse mass 
formed by $\met$ and any of the three leading $b$-tagged jets in the event:
\begin{equation*}
\mtb =  \mathrm{min}_{i\leq 3}  \left ( \sqrt{2\pt^{b\mathrm{-jet}_i}
    \met\{1-\cos[\Delta\phi(\vec{\pt}^{\mathrm{miss}},b\mathrm{-jet}_i )]\}} \right ).
\end{equation*}
It is bounded below the top quark mass for semileptonic $\ttbar$
events while peaking at higher values for Gbb and Gtt events.

The signal regions require either zero or at least one
lepton. The requirement of a signal lepton, with the additional requirements
on jets, $\met$ and event variables described in Section~\ref{sec:signalregion}, render the multijet background
negligible for the $\ge$ 1-lepton signal regions. For the 0-lepton
signal regions, the minimum azimuthal angle between $\vec{\pt}^{\mathrm{miss}}$ and the leading four
small-$R$ jets in the event, $\dphimin$, is required to be greater
than 0.4:
\begin{equation*}
\dphimin = \textrm{min}(|\phi_{\mathrm{jet}_1} -
\phi_{\vec{\pt}^{\mathrm{miss}}}|,...,|\phi_{\mathrm{jet}_4} - \phi_{\vec{\pt}^{\mathrm{miss}}}|) > 0.4.
\end{equation*}
This requirement ensures that the multijet background, which can produce large
$\met$ if containing 
poorly measured jets or neutrinos emitted close to the axis of a jet, is also negligible in the
0-lepton signal regions (along with the other
requirements on jets, $\met$ and event variables described in Section~\ref{sec:signalregion}). 

\begin{figure}[htbp]
\centering
\subfigure[]{\includegraphics[width=0.43\textwidth]{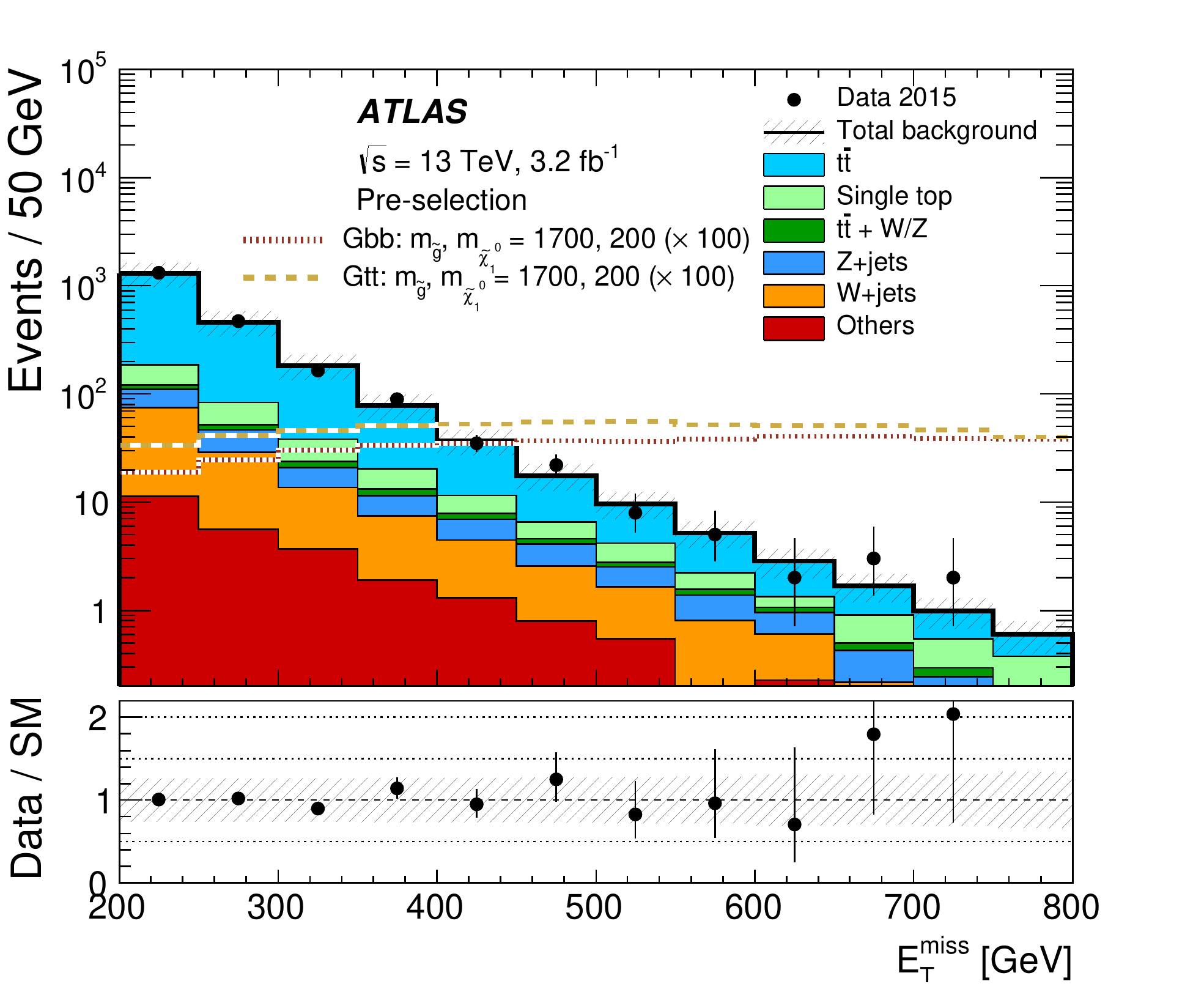}}
\subfigure[]{\includegraphics[width=0.43\textwidth]{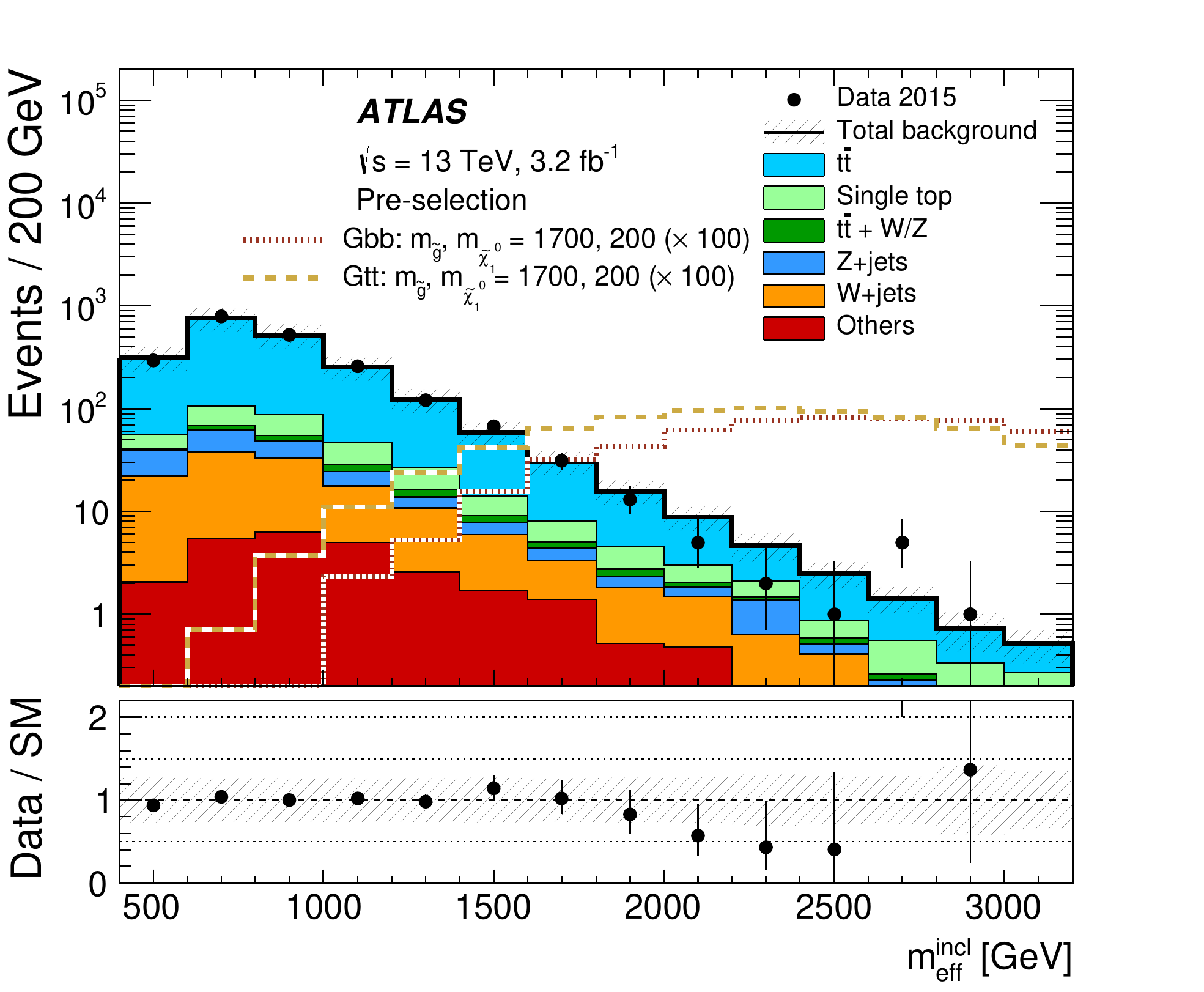}}
\subfigure[]{\includegraphics[width=0.43\textwidth]{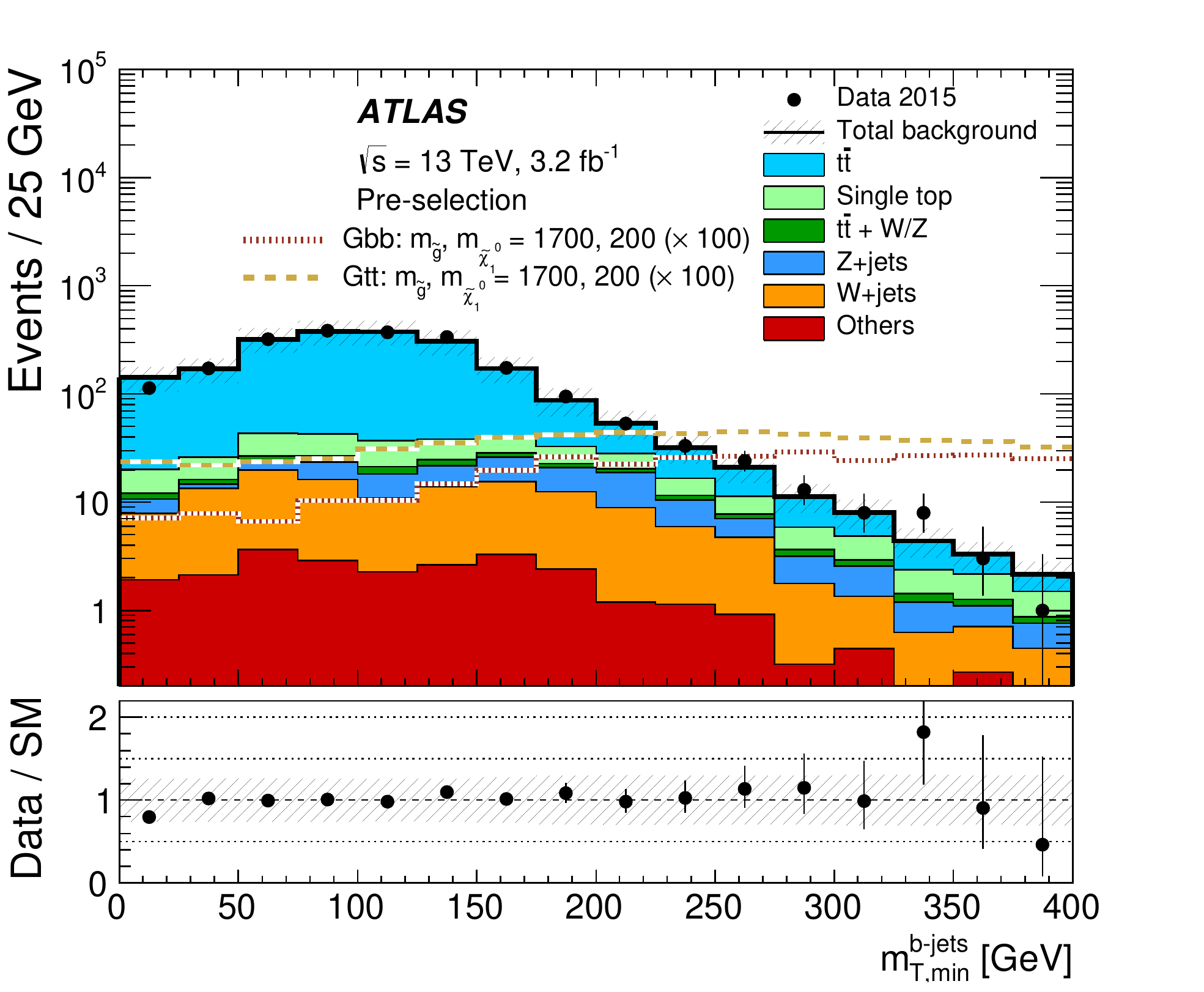}}
\subfigure[]{\includegraphics[width=0.43\textwidth]{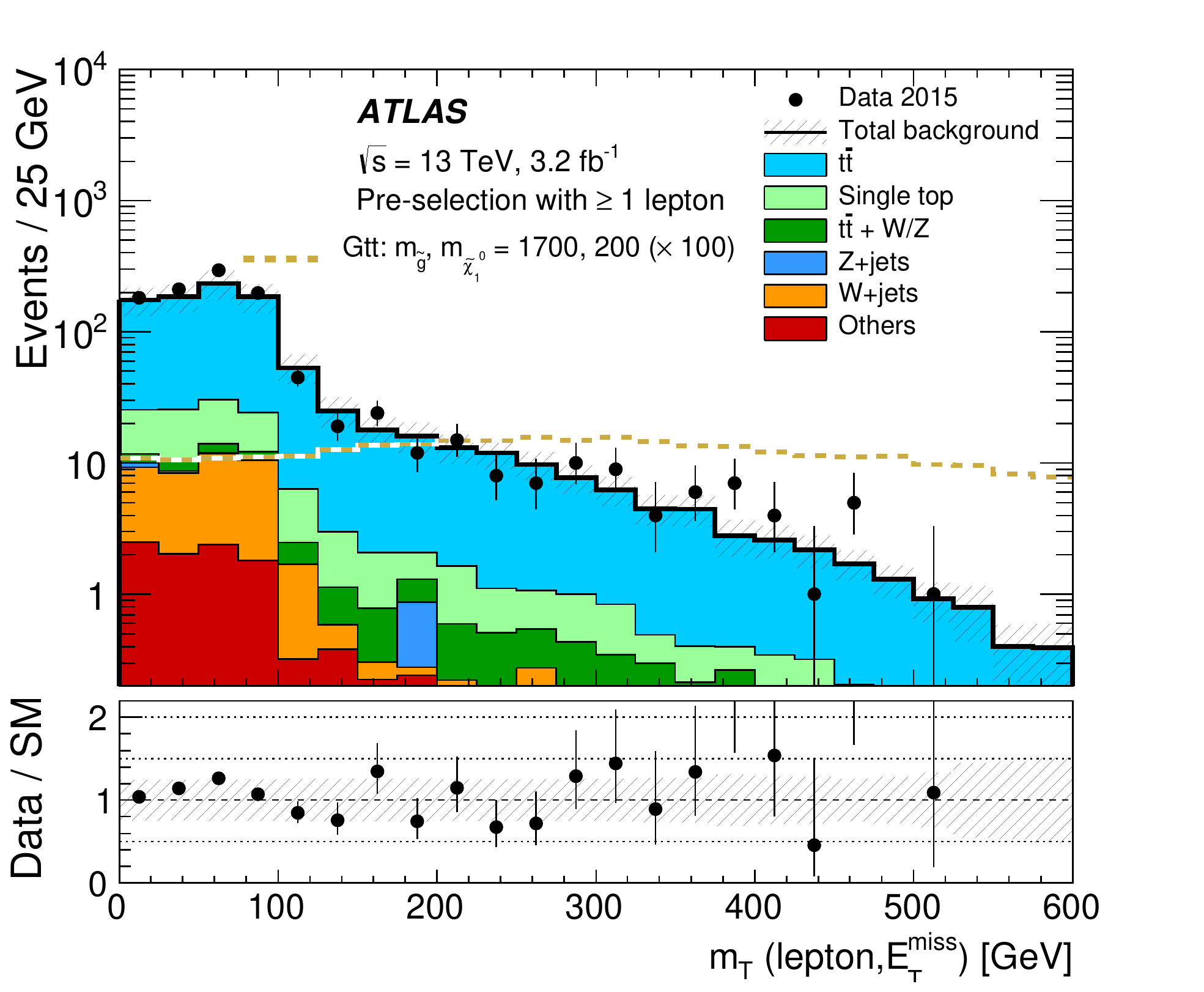}}
\caption{Distributions of kinematic variables  in the
preselection region described in the text: (a) \met, (b)
  \meffi, (c) \mtb, and (d) 
 \mt\ (for preselected events with at least one signal lepton). The statistical and
experimental 
systematic uncertainties are included in the uncertainty band, where
the systematic uncertainties are defined in Section~\ref{sec:syst}.   The lower part of each figure shows
the ratio of data to
          the background prediction. All backgrounds (including
          $\ttbar$) are
          normalized using the best available theoretical calculation
          described in Section~\ref{sec:samples}. The background
          category ``Others'' includes
          $\ttbar h$, $t\bar{t}t\bar{t}$ and diboson events. Example signal models
          with cross-sections enhanced by a factor of 100 are overlaid
          for comparison. 
} \label{fig:presel_var}
\end{figure}
\begin{figure}[htbp]
\centering
\subfigure[]{\includegraphics[width=0.43\textwidth]{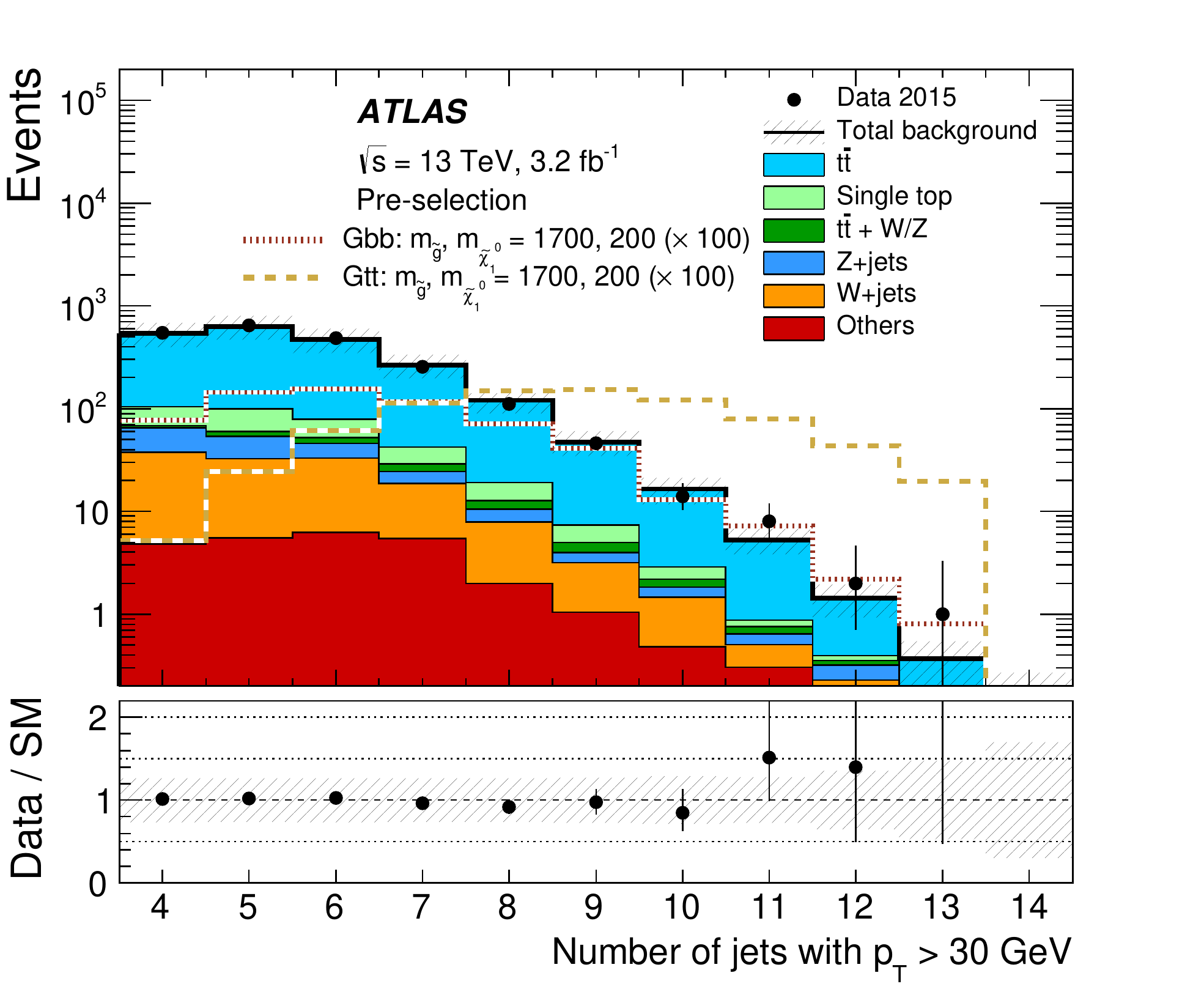}}
\subfigure[]{\includegraphics[width=0.43\textwidth]{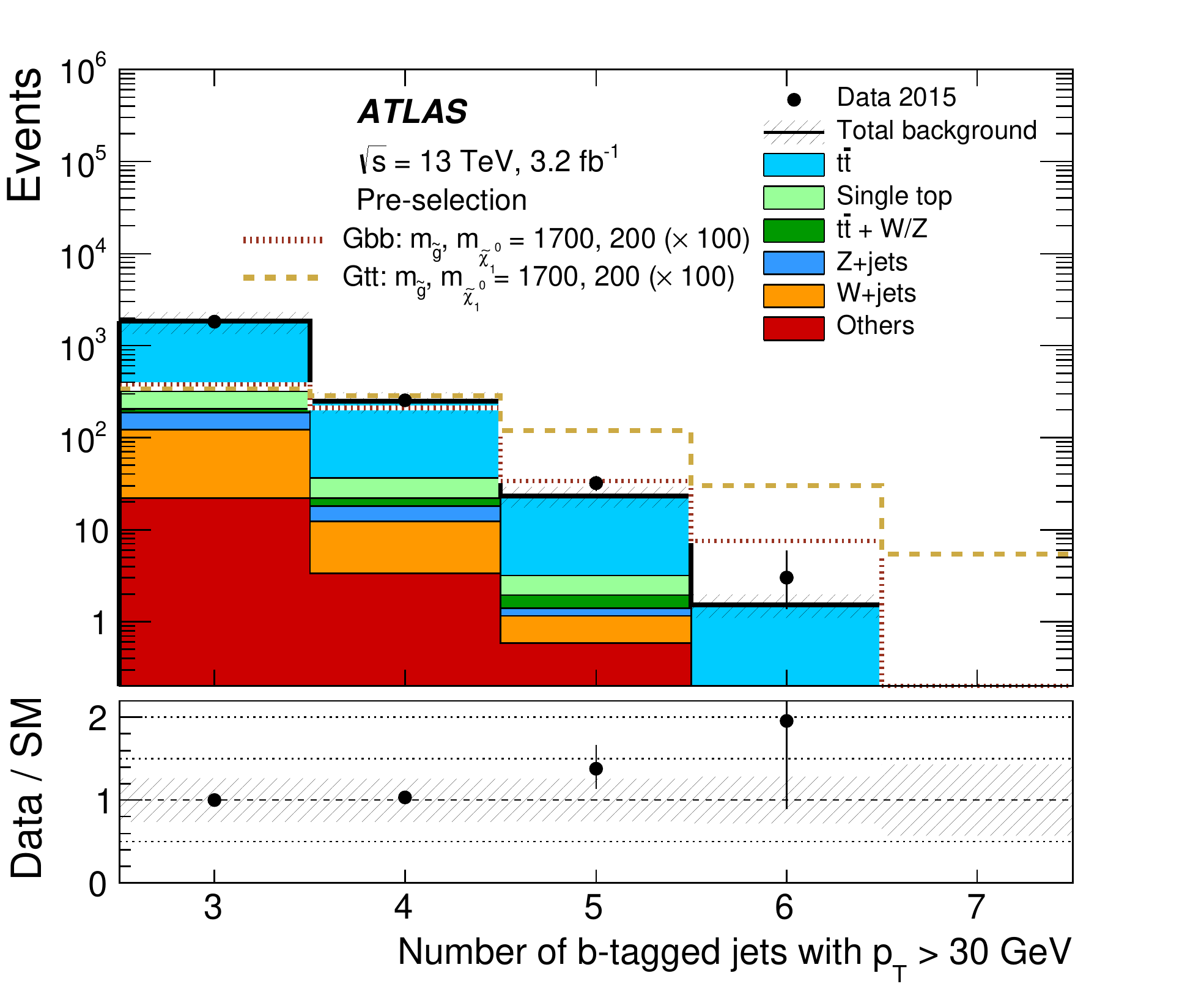}}
\subfigure[]{\includegraphics[width=0.43\textwidth]{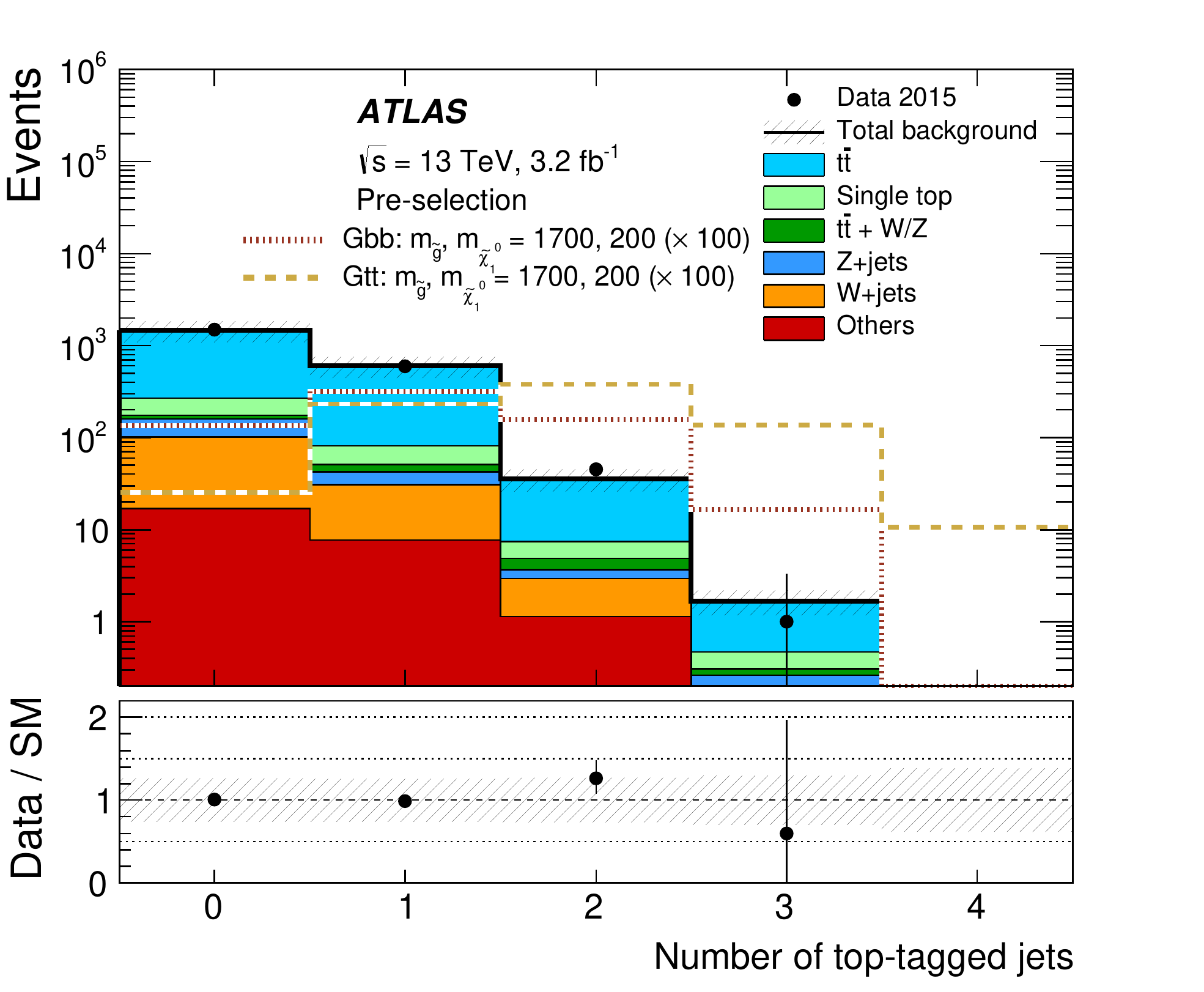}}
\subfigure[]{\includegraphics[width=0.43\textwidth]{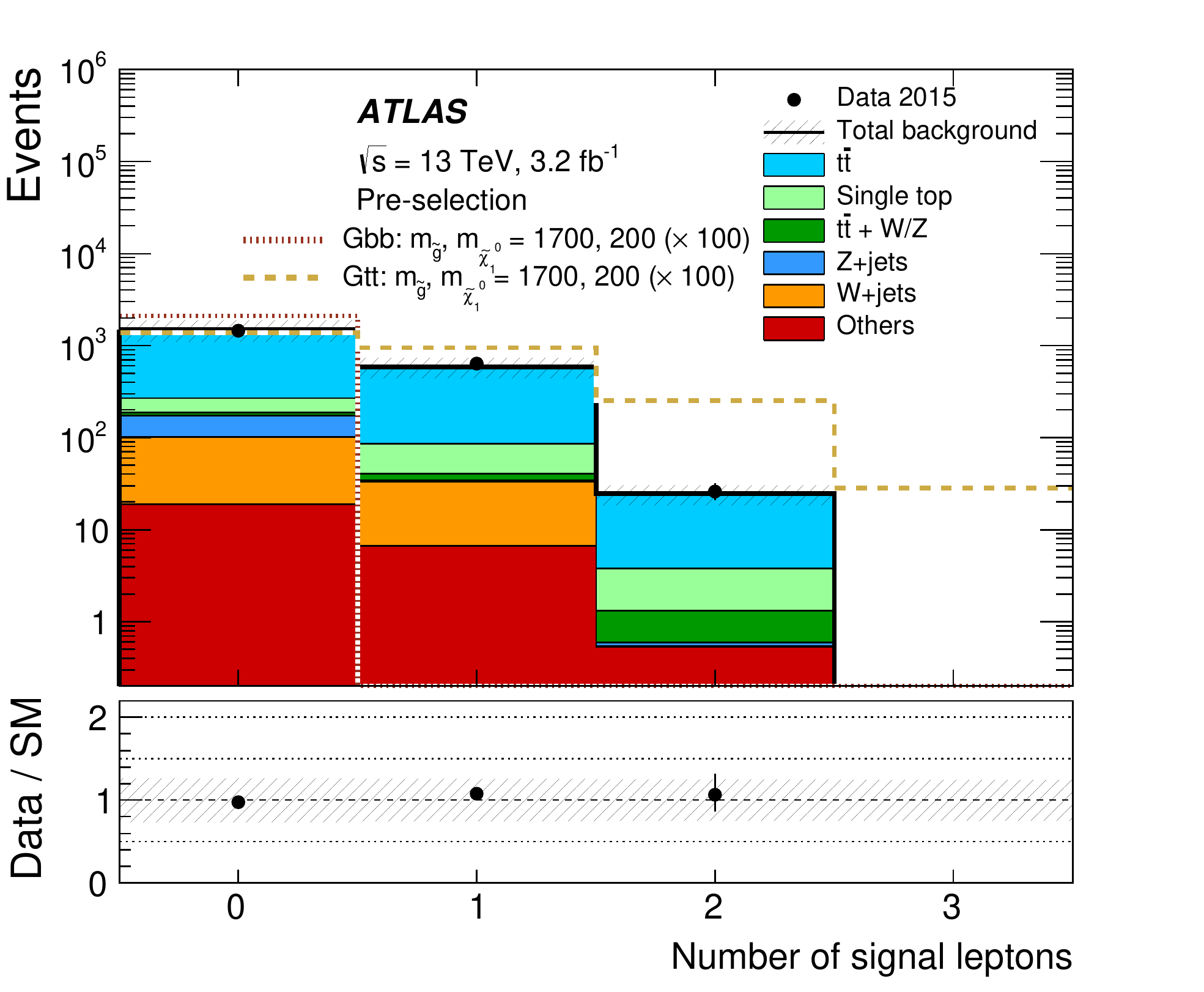}}
\caption{Distributions of the number of: (a) signal jets, (b)
  $b$-tagged jets, (c) top-tagged large-$R$ jets, and (d) 
  signal leptons in the
preselection region described in the text. The statistical and
experimental 
systematic uncertainties are included in the uncertainty band, where
the systematic uncertainties are defined in Section~\ref{sec:syst}.   The lower part of each figure shows
the ratio of data to
          the background prediction. All backgrounds (including
          $\ttbar$) are
          normalized using the best available theoretical calculation
          described in Section~\ref{sec:samples}. The background
          category ``Others'' includes
          $\ttbar h$, $t\bar{t}t\bar{t}$ and diboson events. Example signal models
          with cross-sections enhanced by a factor of 100 are overlaid
          for comparison. 
} \label{fig:presel_n}
\end{figure}

Figure~\ref{fig:presel_var} shows the kinematic distributions of \met,
\meffi, \mtb\ and \mt\ for a preselection that
requires $\met > 200 \gev$, at least four signal jets of which at least
three must be $b$-tagged, and $\dphimin >
0.4$. Figure~\ref{fig:presel_n} shows the multiplicity of signal jets,
$b$-tagged signal jets, top-tagged large-$R$ jets and signal leptons in the
preselection. 
Good agreement between data and simulation is observed.
Example signal models with enhanced cross-sections are overlaid for comparison. 

\subsection{Signal regions}
\label{sec:signalregion}

The signal regions are designed by optimizing the expected
signal discovery reach for the 2015 dataset. They are defined in the
leftmost column of Tables~\ref{tab:GbbEvsel}, \ref{tab:Gtt0LEvsel}
and \ref{tab:Gtt1LEvsel} for the Gbb, Gtt 0-lepton and Gtt 1-lepton
channels, respectively, and are discussed below. These tables also
contain the definition of the control regions used to 
normalize the $\ttbar$ background, discussed in Section~\ref{sec:controlregion},
and the validation regions used to cross-check the background estimate
and which are discussed in Section~\ref{sec:validationregion}. 
The following region  nomenclature is used in the remainder of the paper. Signal, control and
validation region names start with the prefix ``SR'', ``CR'' and
``VR'', respectively, and with the type of validation region specified
for the Gtt validation regions. The name of the region is
completed by the type of model targeted and a letter corresponding
to the level of mass splitting between the gluino and the LSP. For
example the validation region that cross-checks the extrapolation over
$\mt$ for the Gtt 1-lepton region A is denoted by ``VR-$\mt$-Gtt-1L-A''.

The experimental signature for the Gbb model is characterized by four
high-$\pt$ $b$-jets, large $\met$ and no leptons (Figure~\ref{fig:feynman-Gluino-topbottom-offshell}(a)). The following
requirements are applied to all Gbb signal regions. 
Events containing a candidate lepton are vetoed and at least four signal small-$R$ jets
are required, of which at least three must be $b$-tagged. The remaining multijet background is rejected by requiring $\dphimin > 0.4$.
The Gbb signal
regions are described in the leftmost
column of Table~\ref{tab:GbbEvsel}.
The three signal regions A, B and C are designed to cover Gbb models with
large ($\gtrsim 1 \tev$), moderate (between $\approx 200 \gev$ and
$\approx 1 \tev$) and small ($\lesssim 200 \gev$) mass splittings
between the gluino and the
LSP, respectively. All regions feature stringent cuts on
$\met$, $\meffe$ and the jet transverse momentum ${\pt}^\mathrm{jet}$.

 \begin{table}[h]
    \centering
 \renewcommand{\arraystretch}{1.5}
         \begin{tabular}{c c c c c}
        \toprule 
\multicolumn{5}{c}{Criteria common to all Gbb regions: $\geq$ 4 signal
  jets,  $\geq$ 3 $b$-tagged jets} \\ 
        \midrule 
         &  Variable & Signal region  &  Control region & Validation region \\ \midrule
          \multirow{5}{*}{\begin{minipage}{3cm}\centering Criteria common\\ to all
              regions of the same type \end{minipage}} & $N^\mathrm{Candidate\
            Lepton}$ & = 0
          & $-$  & = 0 \\
          &  $N^\mathrm{Signal\ Lepton}$ & $-$ & = 1 &  $-$ \\
          &  $\dphimin$ &$>0.4$ & $-$ & $>0.4$ \\
          & $\mtb$  & $-$ & $-$ & $< 160$ \\
           & $\mt$  & $-$ & $ < 150 $ & $-$ \\
        \midrule 
         \multirow{3}{*}{\begin{minipage}{4cm}\centering Region A\\  (Large mass
             splitting) \end{minipage}} & ${\pt}^\mathrm{jet}$  &
         $>  90$ & $>  90 $ & $ >  90 $ \\
         &  $\met$ & $ > 350 $  & $> 250$   &  $ > 250 $  \\
         & $\meffe$ & $> 1600 $ & $ > 1200 $ & $ <
         1400 $ \\   \midrule
         \multirow{3}{*}{\begin{minipage}{4cm}\centering Region B\\
             (Moderate mass splitting) \end{minipage}} & ${\pt}^\mathrm{jet}$  &
         $>  90$ & $>  90 $ & $ >  90 $ \\
         &  $\met$ & $ > 450 $  & $> 300$   &  $ > 300 $  \\
         & $\meffe$  & $> 1400 $ & $ > 1000 $ & $ <
         1400 $ \\   \midrule
         \multirow{3}{*}{\begin{minipage}{4cm}\centering Region C\\
             (Small mass splitting) \end{minipage}} & ${\pt}^\mathrm{jet}$  &
         $>  30$ & $>  30 $ & $ >  30 $ \\
         &  $\met$ & $ > 500 $  & $> 400$   &  $ > 400 $  \\
         & $\meffe$  & $> 1400 $ & $ > 1200 $ & $ <
         1400 $ \\   \bottomrule
      \end{tabular}   
   \caption{Definitions of the Gbb signal, control and validation
     regions. The unit of all kinematic variables is GeV except
     $\dphimin$, which is in radians. The jet \pt\ requirement is also
     applied to $b$-tagged jets.}     
   \label{tab:GbbEvsel}
 \end{table}

The experimental signature for the Gtt model
is characterized
by several high-$\pt$ jets of which four
are $b$-jets, large $\met$ and potentially leptons (Figure~\ref{fig:feynman-Gluino-topbottom-offshell}(b)). 
The Gtt signal regions are classified into regions with a signal
lepton veto (0-lepton channel) and regions with at least one signal lepton (1-lepton
channel). The Gtt 0-lepton signal regions are defined in the leftmost
column of
Table~\ref{tab:Gtt0LEvsel}.
In all Gtt 0-lepton signal regions at least eight signal jets,
$\dphimin > 0.4$  and $\mtb > 80 \gev$ are
required. Three Gtt 0-lepton signal regions are defined to cover
Gtt models with decreasing mass splitting between the gluino and the
sum of the mass of the two top quarks and the
LSP: A ($\gtrsim 1 \tev$), B (between $\approx 200 \gev$ and
$\approx 1 \tev$) and C ($\lesssim 200 \gev$). In the large and moderate mass splitting scenarios,
the top quarks tend to have a large $\pt$, and at least one top-tagged
large-$R$ jet
is required ($N^\mathrm{top} \ge 1$). 
The requirements on $\met$ and $\meffi$ decrease with the mass splitting
between the gluino and the LSP. However, the required number of $b$-tagged jets
$N^{b\mathrm{-jet}}$ is tightened to four for the lower mass splitting
regions B and C in order to maintain a high background rejection despite the
softer signal kinematics.

The Gtt 1-lepton signal regions are defined in the leftmost
column of
Table~\ref{tab:Gtt1LEvsel}. Two signal regions A and B are
defined to cover Gtt models with decreasing mass difference between
the gluino and the LSP. In all signal regions at least one signal
lepton, at least
six signal jets (${\pt}^\mathrm{jet} >  30 \gev$) and $\mt > 150 \gev$ are required.
Region A has tighter requirements on $\meffi$ ($\meffi > 1100 \gev$) and the
number of top-tagged large-$R$ jets ($N^\mathrm{top} \ge 1$). 
 Region B has a softer
requirement on $\meffi$ than region A, but it features a tighter
cut on $\met$ to achieve a satisfactory
background rejection without requiring a top-tagged large-$R$ jet.
\begin{table}[h]
    \centering
 \renewcommand{\arraystretch}{1.5}
         \begin{tabular}{c c c c c c}
           \toprule
           \multicolumn{6}{c}{Criteria common to all Gtt 0-lepton regions: ${\pt}^\mathrm{jet} >  30 \gev$} \\ 
        \midrule 
          & Variable & Signal region  &  Control region & VR1L & VR0L \\ \midrule
          \multirow{5}{*}{\begin{minipage}{3cm}\centering Criteria common\\ to all
              regions of the same type \end{minipage}} &
          $N^\mathrm{Signal\ Lepton}$ & = 0 & = 1  & = 1   & = 0  \\
          &  $\dphimin$ & $>0.4$ & $-$ & $-$ & $>0.4$ \\
           & $N^\mathrm{jet}$ &  $\ge 8$ & $\ge 7$ & $\ge 7$ & $\ge 8$ \\
          & $\mtb$  & $ > 80 $ & $-$ &  $ > 80 $ & $ < 80 $ \\
          & $\mt$ & $-$ & $ < 150 $ & $ < 150 $ & $-$ \\ \midrule
         \multirow{4}{*}{\begin{minipage}{4cm}\centering Region A\\  (Large mass
             splitting) \end{minipage}} &  $\met$  & $> 400 $  & $ > 250 $   &  $ > 250 $  &  $ > 200 $ \\
         & $\meffi$  & $> 1700 $ & $ > 1350 $ & $ >1350 $ & $ > 1400 $\\
           & $N^{b\mathrm{-tag}}$ & $ \ge 3$ & $\ge 3$ & $ \ge 3$ & $ \ge 2$ \\
           & $N^\mathrm{top} $ & $\ge 1$ & $\ge 1$ & $\ge 1$ & $ \ge 1$ \\  
         \midrule
         \multirow{4}{*}{\begin{minipage}{4cm}\centering Region B\\  (Moderate mass
             splitting) \end{minipage}}  &  $\met$  & $> 350 $  & $ > 200 $   &  $ > 200 $  &  $ > 200 $ \\
         & $\meffi$  & $> 1250 $ & $ > 1000 $ & $ >1000 $ & $ > 1100 $\\
           & $N^{b\mathrm{-tag}}$ & $ \ge 4$ & $\ge 4$ & $ \ge 4$ & $ \ge 3$ \\
           & $N^\mathrm{top} $ & $\ge 1$ & $\ge 1$ & $\ge 1$ & $ \ge  1$ \\  
         \midrule 
         \multirow{3}{*}{\begin{minipage}{4cm}\centering Region C\\  (Small mass
             splitting) \end{minipage}}   &  $\met$  & $> 350 $  & $ > 200 $   &  $ > 200 $  &  $ > 200 $ \\
         & $\meffi$  & $> 1250 $ & $ > 1000 $ & $ >1000 $ & $ > 1250 $\\
           & $N^{b\mathrm{-tag}}$ & $ \ge 4$ & $\ge 4$ & $ \ge 4$ & $ \ge 3$ \\ 
         \bottomrule                    
      \end{tabular}   
   \caption{Definitions of the Gtt 0-lepton signal, control and
     validation regions.  The unit of all kinematic variables is \gev\ except
     $\dphimin$, which is in radians.
     The jet \pt\ requirement is also applied to $b$-tagged jets.}     
   \label{tab:Gtt0LEvsel}
 \end{table}

 \begin{table}[h]
    \centering
 \renewcommand{\arraystretch}{1.5}
         \begin{tabular}{c c c c c c}
        \toprule 
\multicolumn{6}{c}{Criteria common to all Gtt 1-lepton regions: $\ge
  1$ signal lepton, ${\pt}^\mathrm{jet} >  30 \gev$} \\ 
        \midrule            
          & Variable & Signal region  &  Control region & VR-$\mt$ & VR-$\mtb$ \\ \midrule
          \multirow{3}{*}{\begin{minipage}{3cm}\centering Criteria common\\ to all
              regions of the same type \end{minipage}} 
                                                     & $\mt$  & $> 150$ & $ < 150 $  &  $ > 150 $ & $ < 150 $ \\
                                                     & $N^\mathrm{jet}$&  $\ge 6$ &  $ = 6$ &  $\ge 5$ &  $\ge 6$ \\
                                                     &
                                                     $N^{b\mathrm{-tag}}$
                                                     & $\ge 3$ & $\ge 3$ & $ = 3$ & $ = 3$ \\
          \midrule
         \multirow{4}{*}{\begin{minipage}{3cm}\centering Region A\\  (Large mass
             splitting) \end{minipage}} &  $\met$  & $ > 200 $  & $ > 200 $   &  $ > 200 $  &  $ > 200 $ \\
                                                     & $\meffi$  & $ > 1100  $ & $ >1100 $ & $ > 600 $ &  $ > 600  $\\
                                                     & $\mtb$  & $ > 160 $ & $-$ &  $ < 160 $ & $ > 140 $ \\
                                                     &
                                                     $N^\mathrm{top}$
                                                     & $ \ge 1$ & $ \ge 1$ & $ \ge 1$ & $ \ge 1$ \\  
         \midrule
         \multirow{3}{*}{\begin{minipage}{3cm}\centering Region B\\
             (Moderate to small mass
             splitting) \end{minipage}} &  $\met$  & $ > 300 $  & $ > 300 $   &  $ > 200 $  &  $ > 200 $ \\
                                                     & $\meffi$ 
                                                     & $ > 900  $ & $  >900 $ & $ > 600 $ &  $ > 600 $\\
                                                     & $\mtb$  & $ > 160 $ & $-$ &  $ < 160 $ & $ > 160 $ \\  
         \bottomrule    
      \end{tabular}   
   \caption{Definitions of the Gtt 1-lepton signal, control and
     validation regions.  The unit of all kinematic variables is \gev.
     The jet \pt\ requirement is also applied to $b$-tagged jets.}     
   \label{tab:Gtt1LEvsel}
 \end{table}

\subsection{Background estimation and $\ttbar$ control regions}
\label{sec:controlregion}

The largest background in all signal regions is $\ttbar$ produced with
additional high-$\pt$ jets. The other relevant backgrounds are $\ttbar
 W$, $\ttbar  Z$, $t\bar{t}t\bar{t}$, $\ttbar h$, single-top, $W$+jets,
 $Z$+jets and diboson events. All of these smaller backgrounds are estimated
 with the simulated event samples normalized to the best available theory calculations described
 in Section~\ref{sec:samples}. The multijet background is estimated
 to be negligible in all regions.

For each signal region, the $\ttbar$ background is normalized in a dedicated
control region.
The $\ttbar$ normalization factor required for the total
predicted yield to match the data in the control region is used
to normalize the $\ttbar$ background in the signal region.
The control regions are designed to be
dominated by $\ttbar$ events and to have negligible signal contamination, while
being kinematically as close as possible to the corresponding signal
region. The latter requirement minimizes the systematic uncertainties
associated with extrapolating the normalization factors from the
control to the signal regions.

The definitions of the control regions are shown next to 
the signal regions in Tables~\ref{tab:GbbEvsel}, \ref{tab:Gtt0LEvsel}
and \ref{tab:Gtt1LEvsel} for the Gbb, Gtt 0-lepton and Gtt 1-lepton
channels, respectively. In both the Gbb and Gtt 0-lepton channels,
exactly one signal lepton is required. This is motivated by 
background composition studies using simulated events which show
that semileptonic $\ttbar$ events, for which the lepton is outside the 
acceptance or is a hadronically decaying $\tau$-lepton, dominate the $\ttbar$
yield in the signal regions. An upper cut on $\mt$ is then applied to
ensure orthogonality with the Gtt 1-lepton signal regions and to 
suppress signal contamination. The jet multiplicity requirement is
reduced to seven jets in the Gtt 0-lepton control regions (from
eight jets in the signal regions), to accept more events and to obtain
a number of jets from top quark decay and parton
shower similar to that in the signal region. Approximately 40--60\%  of
the signal region events contain a hadronically decaying $tau$-lepton that 
is counted as a jet. Orthogonality between Gtt 0-lepton and Gtt 1-lepton 
control regions is ensured by requiring exactly six jets in the Gtt 
1-lepton control regions (as opposed to the requirement of {\em at least} six jets 
in the signal regions). For all Gbb and Gtt 0-lepton control regions, the number of
$b$-tagged jets and top-tagged large-$R$ jets is consistent with the signal
region. The requirements on $\met$ and $\meff$ are, however, relaxed
in the control regions
to achieve a sufficiently large $\ttbar$ yield
and small signal contamination ($\lesssim$ 15\%). 
The Gtt 1-lepton control regions are defined by inverting
the $\mt$ cut and removing the $\mtb$ requirement.  All other
requirements are exactly the same as for the signal regions.

\subsection{Validation regions}
\label{sec:validationregion}

Validation regions are defined to cross-check the background
prediction in regions that are kinematically close to the signal regions 
but yet have a small signal contamination.
They are designed primarily to
cross-check the assumption that the $\ttbar$ normalization extracted
from the control regions can be accurately extrapolated to the signal regions.
Their requirements are shown in the rightmost column(s) of Tables~\ref{tab:GbbEvsel}, \ref{tab:Gtt0LEvsel}
and \ref{tab:Gtt1LEvsel} for the Gbb, Gtt 0-lepton and Gtt 1-lepton
channels, respectively.  Their signal
contamination is less than approximately 30\% for the majority of Gbb
and Gtt model points not
excluded in Run~1.

One validation region per signal region is defined for the Gbb
model. They feature the same requirements as their corresponding
signal region except that upper cuts are applied on $\mtb$ and $\meffe$ to reduce
signal contamination and ensure orthogonality with the signal regions. In addition the requirement on $\met$ is
relaxed to obtain a sufficient $\ttbar$ yield.

For the Gtt 0-lepton channel, two validation regions per
signal region are defined, one requiring exactly one signal lepton
(VR1L) and one with a signal lepton veto (VR0L). The regions VR1L have
exactly the same criteria as their corresponding control regions except
that they require $\mtb > 80 \gev$, similarly to the signal regions,
in order to test the extrapolation over $\mtb$ between the control and
the signal regions. 
Simulation studies show that the heavy-flavor
fraction of the additional jets in the $\ttbar$+jets events
(i.e. $\ttbar + \bbbar$ and $\ttbar + \ccbar$), which suffers from large
theoretical uncertainties, is similar in the signal, control and
VR1L regions. This is achieved by requiring the same number of
$b$-tagged jets for all three types of regions.

While the theoretical uncertainties in the heavy-flavor
fraction of the additional jets in the $\ttbar$+jets events
(i.e. $\ttbar + \bbbar$ and $\ttbar + \ccbar$) are large, they affect
signal, control and the 1-lepton validation regions in a similar way, and are thus largely canceled in
the semi-data-driven $\ttbar$ normalization based on the observed control region yields. 

The VR0L regions have similar requirements on
their corresponding signal regions except that the requirements 
on $\met$, $\meffi$ and the number of $b$-tagged jets are loosened to achieve
sufficient event yields. Furthermore, the criterion $\mtb < 80 \gev$  is
applied to all VR0L regions to ensure orthogonality with the signal
regions. The regions VR0L test the extrapolation of the $\ttbar$ normalization from a
1-lepton to a 0-lepton region. Simulation
studies show that the VR0L regions have a composition of
semileptonic $\ttbar$ events (in particular of hadronically decaying $\tau$-leptons) similar to that in the 
signal regions, while the control and VR1L regions are by construction
dominated by semileptonic $\ttbar$ events with
a muon or an electron. 

Two requirements are different between Gtt 1-lepton control regions
and their corresponding signal regions: the requirement on $\mtb$ (absent in
the control regions) and the requirement on $\mt$ (inverted in the control regions).
Therefore, two validation regions per signal region are defined for the Gtt
1-lepton channel, VR-$\mt$ and VR-$\mtb$, which respectively test, one
at a time, the
extrapolations over $\mt$ and $\mtb$. Exactly three
$b$-tagged jets are required for all 1-lepton validation regions to limit the
signal contamination and to be close to the signal regions. For the VR-$\mt$ regions, the same requirement
$\mt > 150 \gev$ as in the signal region is applied but the criterion on $\mtb$
is inverted. Other requirements
are relaxed to achieve sufficiently large background yields and  small signal 
contamination. For the VR-$\mtb$ regions, the signal region requirement on $\mtb$ is applied
(slightly loosened to 140~\gev\ instead of 160~\gev\ in region A) and the
criterion on $\mt$ is inverted. Again, other requirements are generally
relaxed. Simulation studies show that $\ttbar$ dilepton events
dominate in the signal regions, in particular due to the requirement on $\mt$, while semileptonic $\ttbar$ events
dominate in the control regions. This extrapolation is cross-checked
by the VR-$\mt$ regions, which have a $\ttbar$ dileptonic
fraction similar to that in the signal regions. 

\section{Systematic uncertainties}
\label{sec:syst}
The largest sources of detector-related systematic uncertainties in this analysis relate to 
the jet energy scale (JES), jet energy resolution (JER) and the $b$-tagging efficiencies and 
mistagging rates. The JES uncertainties are obtained by extrapolating the uncertainties derived 
from $\sqrt{s}= 8 \tev$ data and simulations to $\sqrt{s}= 13 \tev$~\cite{Run2-JES}. The 
uncertainties in the energy scale of the small-$R$ jets are propagated to the re-clustered 
large-$R$ jets, which use them as inputs. The JES uncertainties are especially important in the Gtt signal 
regions, since these regions require high jet multiplicities. The impact of these uncertainties on 
the expected background yields in these regions is between 10\% and 25\%. Uncertainties in the JER 
are similarly derived from dijet asymmetry measurements in Run 1 data and extrapolated 
to $\sqrt{s}= 13 \tev$. The impact of the JER uncertainties on the background yields are in the range of 1--10\%.

Uncertainties in the measured $b$-tagging efficiencies and mistagging rates are the subleading sources of 
experimental uncertainties in the Gtt 1-lepton signal regions and the leading source in the Gtt 0-lepton 
and Gbb regions. Uncertainties measured in $\sqrt{s}= 8 \tev$ data are extrapolated to $\sqrt{s}= 13 \tev$
, with the addition of the new IBL system in Run 2 taken into
account. Uncertainties for jet \pt above 300~\gev\ are estimated using simulated events. The impact of the $b$-tagging uncertainties on the expected background 
yields in the Gbb and Gtt 0-lepton signal regions is around 22--30\%, and around 15\% in the Gtt 1-lepton signal regions.

The uncertainties associated with lepton reconstruction and energy measurements have very small impact 
on the final results. All lepton and jet measurement uncertainties are propagated to the calculation 
of \met, and additional uncertainties are included in the scale and resolution of the soft term. The 
overall impact of the \met soft term uncertainties is also small.

Uncertainties in the modeling of the \ttbar background are evaluated using additional samples varied 
by each systematic uncertainty. Hadronization and parton showering uncertainties are estimated using 
a sample generated with \POWHEG  and showered by \MYHERWIG++ v2.7.1~\cite{Bahr:2008pv} with the UEEE5 
underlying-event tune~\cite{ueee5}. Systematic uncertainties in the modeling of initial- and final-state 
radiation are explored with two alternative settings of \POWHEG, both of which are showered by \PYTHIA 
v6.428 as for the nominal sample. The first of these uses the PERUGIA2012radHi tune and has the renormalization 
and factorization scales set to twice the nominal value, resulting in more radiation in the final state. 
It also has $h_\mathrm{damp}$ set to 2$m_\mathrm{top}$. The second sample, using the PERUGIA2012radLo tune, 
has $h_\mathrm{damp}=m_\mathrm{top}$ and the renormalization and factorization scales are set to half of 
their nominal values, resulting in less radiation in the event. In each case, the uncertainty is taken as 
the deviation in the expected yield of \ttbar background with respect to the nominal sample. The uncertainty 
due to the choice of generator is estimated by comparing the expected yields obtained using a \ttbar sample 
generated with \MGMCatNLO~, and one that is generated with \POWHEG. Both of these samples are showered with 
\MYHERWIG++ v2.7.1. Finally, a 30\% uncertainty is assigned to the cross-section of \ttbar events with additional 
heavy-flavor jets in the final state, in accordance with the results of the ATLAS measurement of this 
cross-section at $\sqrt{s}= 8 \tev$~\cite{ttbb-xsec}. Uncertainties in single-top and $W/Z+$jets background 
processes are similarly estimated by comparisons between the nominal sample and samples with different generators, 
showering models and radiation tunes. An additional 5\% uncertainty is included in the cross-section of 
single-top processes~\cite{Kant:2014oha}. A 50\% constant uncertainty is assigned to each of the remaining small backgrounds.
The variations in the expected background yields due to \ttbar modeling uncertainties 
range between 10\% and 30\% for the Gbb signal regions, and between 47\% and 57\% in most Gtt signal regions. 
The impact of the modeling uncertainties for the smaller backgrounds on these yields is consistently below 10\% 
in all signal regions. The uncertainties in the cross-sections of signal processes are determined from an envelope of different 
cross-section predictions, as described in Section~\ref{sec:samples}. 

The cumulative impact of the systematic uncertainties listed above on the background yields ranges between 23\% and 63\%, 
depending on the signal region. The typical impact on the signal yields is in the range 10-30\%.


\section{Results}
\label{sec:result}
The SM background expectation  is determined separately in each signal region with a profile likelihood fit
\cite{HFpaper}, referred to as a background-only fit. The fit uses as a
constraint the observed event yield in the associated control region
to adjust the $\ttbar$ normalization, assuming that a signal does
not contribute to this yield, and applies that normalization factor to
the number $\ttbar$ events predicted by simulation in the signal region. The numbers of observed and predicted
events in each control region are described by Poisson probability
density functions.  
The systematic uncertainties in
the expected values are included in the fit as nuisance
parameters. They are constrained by Gaussian distributions with
widths corresponding to the sizes of the uncertainties and 
are treated as correlated, when appropriate, between the various regions. 
The product of the 
various probability density functions forms the likelihood, which the
fit maximizes by adjusting the $\ttbar$ normalization and the nuisance
parameters. The inputs to the fit for each signal region are the number
of events observed in its associated control region and the
number of events predicted by simulation in each region for 
all background processes.  

Figure~\ref{fig:pullVR} shows the results of the background-only fit to the control regions, extrapolated to the validation regions. 
The number of events predicted by the background-only fit is compared to
the data in the upper panel. 
The pull, defined by the difference
between the observed number of events ($n_\mathrm{obs}$) and the predicted
background yield ($n_\mathrm{pred}$)
divided by the total uncertainty ($\sigma_\mathrm{tot}$), is shown for each region in the lower panel. No
evidence of significant background mismodeling is observed in the
validation regions. 
There is a certain tendency for the predicted
background to be above the data, 
in particular for the Gtt-0L validation regions, 
but the results in the validations regions of a given channel are not
independent.  The validation and control regions of different mass
splittings can overlap, with the overlap fraction ranging from
approximately 30\% to 70\% for Gtt-0L. Furthermore, the uncertainties in the predicted
yield are dominated by the same
(correlated) systematic uncertainties. 

\begin{figure}[htbp]
\centering
\includegraphics[width=1.0\textwidth]{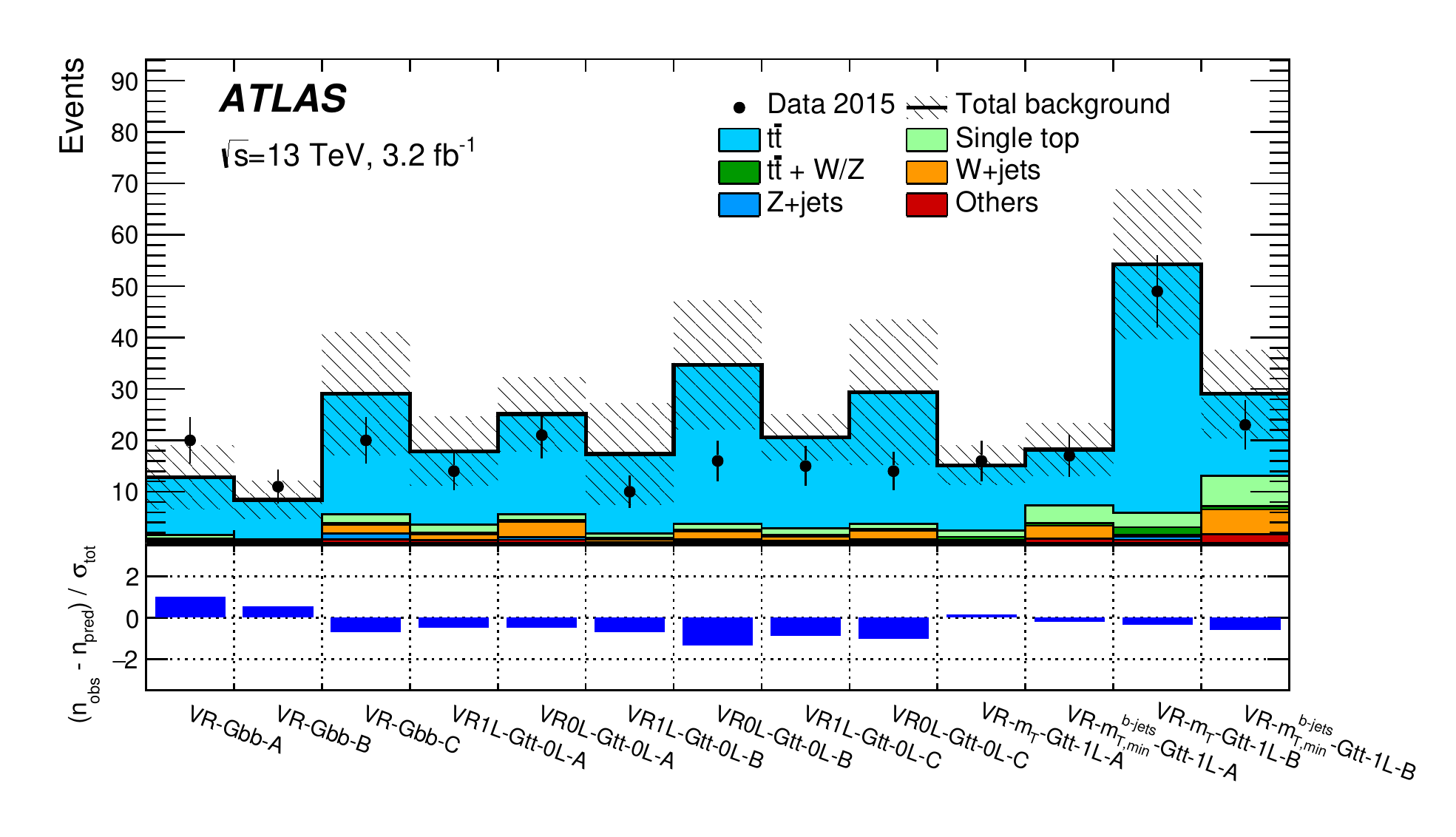}
\caption{Results of the likelihood fit extrapolated to the validation
  regions. The $\ttbar$ normalization is obtained from the fit to the
  control regions. The upper panel shows the observed number of
  events and the predicted background yield. The background
          category ``Others'' includes
          $\ttbar h$, $\ttbar\ttbar$ and diboson events. The lower panel shows the pulls in each validation region.
} \label{fig:pullVR}
\end{figure}

Tables~\ref{tab:yield_Gbb}, \ref{tab:yield_Gtt0L} and \ref{tab:yield_Gtt1L} 
show the observed number of events and predicted number of
background events from the background-only fit in the 
Gbb, Gtt 0-lepton and Gtt 1-lepton signal regions, respectively. 
In addition, the tables show the numbers of signal events expected for some 
example values of gluino and LSP masses in the Gtt and Gbb models.
The event yields in the signal regions are also shown in
Figure~\ref{fig:pullSR}, where the pull is shown for each region in
the lower panel. 
No excess is found above the predicted background. 
The background is dominated by $\ttbar$ events in all
Gbb and Gtt signal regions. The subdominant contributions in the Gbb
and Gtt 0-lepton signal regions are $Z(\to \nu\nu)$+jets and $W(\to \ell
\nu)$+jets events, where for $W$+jets events the lepton is
a nonidentified electron or muon or is a hadronically decaying $\tau$-lepton. In the Gtt
1-lepton signal regions, the subdominant backgrounds are single-top,
$\ttbar W$ and $\ttbar Z$. 

\begin{table*}[htp]
\centering
\begin{tabular}{lccc}
\toprule
        & SR-Gbb-A            &   SR-Gbb-B           &   SR-Gbb-C         \\[-0.05cm]
\midrule
Observed events              & $ 0 $    & $ 0 $    & $ 5 $   \\
\midrule
Fitted background events           &   $ 1.3 \pm 0.4 $ & $ 1.5 \pm 0.6 $  & $ 7.6 \pm 1.7 $  \\
\midrule
\ttbar\            & $ 0.63 \pm 0.30 $ & $ 0.9 \pm 0.5$  & $ 4.3 \pm 1.5 $ \\
$Z$+jets            & $ 0.23 \pm 0.08  $ & $ 0.23 \pm 0.09 $  & $ 1.2 \pm 0.5 $ \\
$W$+jets            & $ 0.17 \pm 0.06 $ & $ 0.13 \pm 0.05 $  & $ 0.82 \pm 0.28 $ \\
Single-top            & $ 0.25 \pm 0.14 $ & $ 0.15 \pm 0.14 $  & $0.65 \pm 0.33 $ \\
$\ttbar W/Z$            & $ < 0.1 $ & $ < 0.1 $  & $ 0.22 \pm 0.12 $ \\
Others           & $ < 0.1 $ & $ < 0.1 $  & $ 0.39 \pm 0.22 $ \\   
 \midrule
 MC-only background prediction & $ 1.7 $ & $ 1.5 $ & $ 6.7 $ \\  
\midrule
$\mu_{\ttbar}$ & $ 0.64 \pm 0.33 $ & $ 1.0 \pm 0.4 $ & $ 1.2 \pm 0.4 $ \\ 
\midrule
Gbb (\mglu=1700 \gev, \mchi=200 \gev) & $ 3.8 $ & $ 3.5 $ & $ 4.2 $ \\  
Gbb (\mglu=1400 \gev, \mchi=800 \gev) & $ 5.3 $ & $ 7.2 $ & $ 10.5 $ \\  
\bottomrule
\end{tabular}

\caption{Results of the likelihood fit extrapolated to the Gbb signal regions. The
  uncertainties shown include all systematic uncertainties. The data in the
    signal regions are not included in the fit. 
The row ``MC-only background prediction'' provides the total background prediction when the
$\ttbar$ normalization is obtained from a theoretical
calculation~\cite{Czakon:2011xx}. The $\ttbar$ normalization factor $\mu_{\ttbar}$
obtained from the corresponding $\ttbar$ control region is also provided.
The background category ``Others'' includes $\ttbar h$, $\ttbar \ttbar$
  and diboson events. Expected yields for two example Gbb models are also shown. }
\label{tab:yield_Gbb}
\end{table*}

\begin{table*}[htp]
\centering
\begin{tabular}{lccc}
\toprule
        & SR-Gtt-0L-A            &   SR-Gtt-0L-B           &   SR-Gtt-0L-C         \\[-0.05cm]
\midrule
Observed events              & $ 1 $    & $ 1 $    & $ 1 $   \\
\midrule
Fitted background events           &   $ 2.1 \pm 0.5 $ & $ 2.9 \pm 1.8 $  & $ 3.4 \pm 1.8 $  \\
\midrule
\ttbar\            & $ 1.4 \pm 0.4 $ & $ 2.4 \pm 1.7 $  & $ 2.6 \pm 1.8 $ \\
$Z$+jets            & $ 0.22 \pm 0.09 $ & $ 0.11 \pm 0.06$  & $ 0.14 \pm 0.07 $ \\
$W$+jets            & $ 0.19 \pm 0.08 $ & $ 0.14 \pm 0.06 $  & $ 0.18 \pm 0.08$ \\
Single-top            & $ 0.17 \pm 0.17 $ & $ 0.14 \pm 0.13 $  & $ 0.17 \pm 0.15 $ \\
$\ttbar W/Z$              & $ < 0.1 $ & $ < 0.1 $  & $ 0.10 \pm 0.07 $ \\
Others           &  $ < 0.1 $ &  $ < 0.1 $  & $ 0.20 \pm 0.17 $ \\   
 \midrule
 MC-only background prediction & $ 1.8$ & $ 1.9 $ & $ 2.5 $ \\  
\midrule
$\mu_{\ttbar}$ & $ 1.3 \pm 0.4$ & $ 1.8 \pm 0.8 $ & $ 1.5 \pm 0.7 $ \\ 
\midrule
Gtt (\mglu=1600 \gev, \mchi=200 \gev) & $ 3.8 $ & $ 2.8 $ & $ 2.9 $ \\  
Gtt (\mglu=1400 \gev, \mchi=800 \gev) & $ 2.0 $ & $ 3.7 $ & $ 4.1 $ \\  

\bottomrule

\end{tabular}
\caption{Results of the likelihood fit extrapolated to the Gtt 0-lepton signal regions. The
  uncertainties shown include all systematic uncertainties. The data in the
  signal regions are not included in the fit. 
The row ``MC-only background prediction'' provides the total background prediction when the
$\ttbar$ normalization is obtained from a theoretical
calculation~\cite{Czakon:2011xx}. The $\ttbar$ normalization factor $\mu_{\ttbar}$
obtained from the corresponding $\ttbar$ control region is also provided.
 The category ``Others'' includes $\ttbar h$, $\ttbar \ttbar$
  and diboson events. Expected yields for two example Gtt models are also shown. }

\label{tab:yield_Gtt0L}
\end{table*}

\begin{table*}[htp]
\centering
\begin{tabular}{lcc}
\toprule
       & SR-Gtt-1L-A            &   SR-Gtt-1L-B               \\[-0.05cm]
\midrule
Observed events              & $ 2 $    & $0  $   \\
\midrule
Fitted background events           &   $ 1.2 \pm 0.6 $ & $ 1.2 \pm 0.8 $  \\
\midrule
\ttbar\            & $ 0.8 \pm 0.6 $ & $ 0.8 \pm 0.7 $  \\
$Z$+jets            & -- & $ < 0.1 $   \\
$W$+jets            & $ < 0.1 $ & $ < 0.1 $  \\
Single-top            & $ 0.18 \pm 0.14 $ & $ 0.14 \pm 0.12 $  \\
$\ttbar W/Z$         & $ 0.14 \pm 0.08 $ & $ 0.15 \pm 0.09 $  \\
Others           & $ < 0.1 $ & $ < 0.1 $ \\   
 \midrule
MC-only background prediction &  $ 1.3 $  & $ 1.2 $ \\  
\midrule
$\mu_{\ttbar}$ & $ 0.86 \pm 0.28 $ & $ 1.0 \pm 0.4 $ \\ 
\midrule
Gtt (\mglu=1600 \gev, \mchi=200 \gev) & $ 3.4 $ & $ 3.3 $ \\  
Gtt (\mglu=1400 \gev, \mchi=800 \gev) & $ 3.8 $ & $ 4.6 $ \\  
\bottomrule
\end{tabular}

\caption{Results of the likelihood fit extrapolated to the Gtt 1-lepton signal regions. The
  uncertainties shown include all systematic uncertainties. The data in the
  signal regions are not included in the fit. 
The row ``MC-only background prediction'' provides the total background prediction when the
$\ttbar$ normalization is obtained from a theoretical
calculation~\cite{Czakon:2011xx}. The $\ttbar$ normalization factor $\mu_{\ttbar}$
obtained from the corresponding $\ttbar$ control region is also provided.
 The category ``Others'' includes $\ttbar h$, $\ttbar \ttbar$
  and diboson events. Expected yields for two example Gtt models are also shown. }

\label{tab:yield_Gtt1L}
\end{table*}

\begin{figure}[htbp]
\centering
\includegraphics[width=1.0\textwidth]{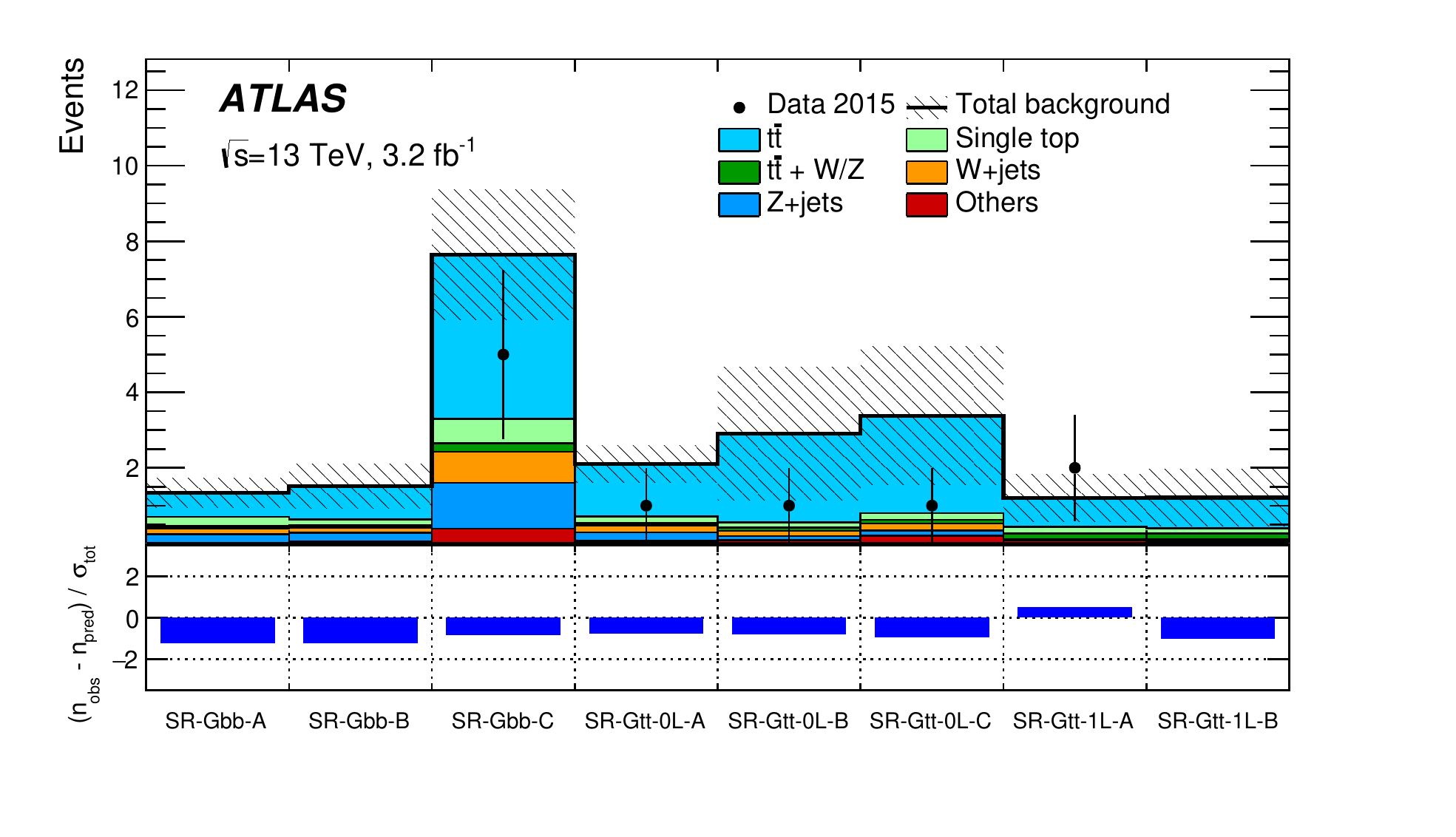}
\caption{Results of the likelihood fit extrapolated to the signal
  regions. The data in the
  signal regions are not included in the fit.  The upper panel shows the observed number of
  events and the predicted background yield. The background
          category ``Others'' includes
          $\ttbar h$, $\ttbar\ttbar$  and diboson events.  The lower panel shows the pulls in each signal region.
} \label{fig:pullSR}
\end{figure}

Figure~\ref{fig:SRplots} shows the $\met$ distributions in
data and simulated samples
for SR-Gbb-B, SR-Gtt-0L-C and SR-Gtt-1L-A, after relaxing the $\met$ threshold to 200~\gev.

\begin{figure}[htbp]
\centering
\subfigure[]{\includegraphics[width=0.45\textwidth]{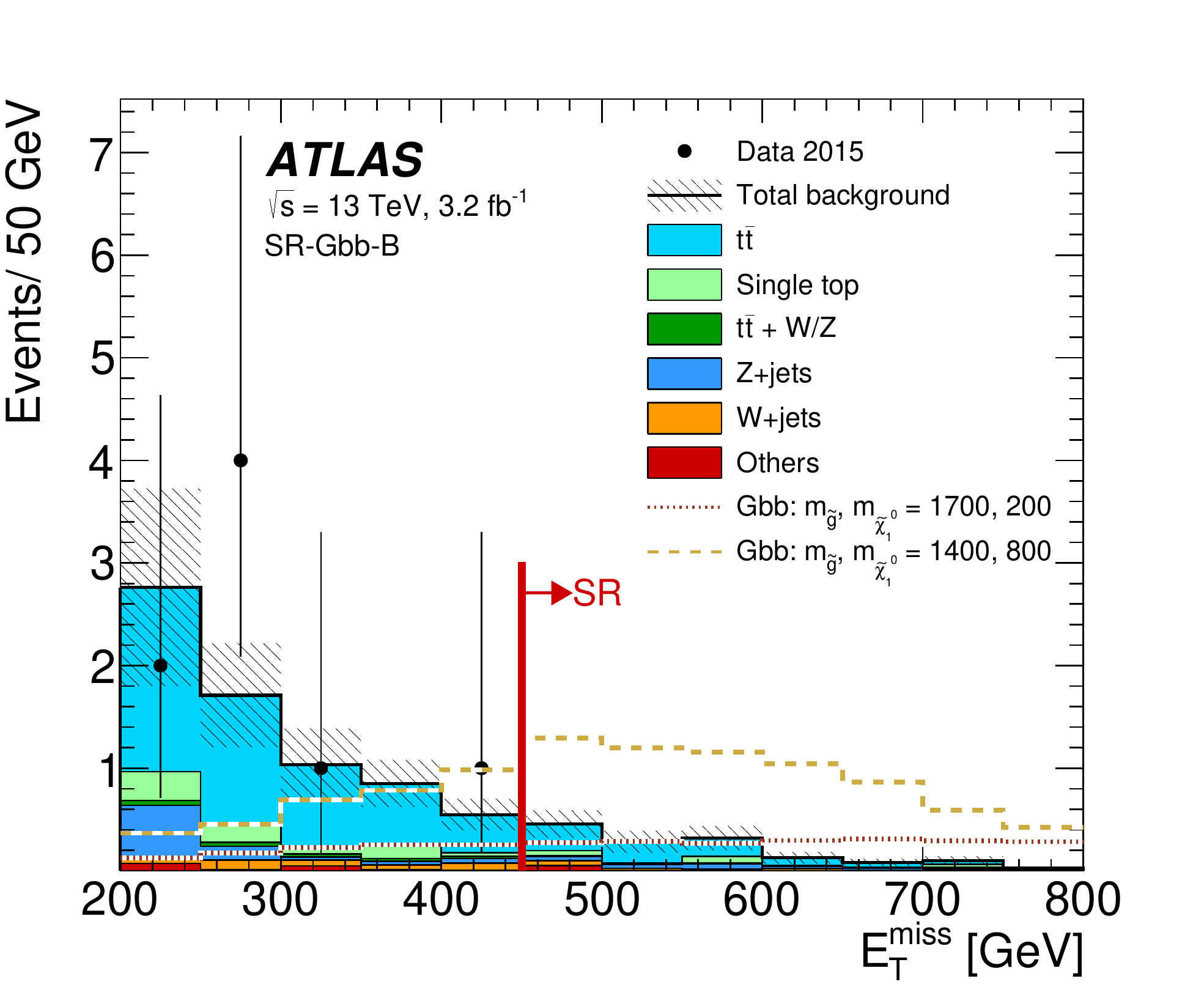}}
\subfigure[]{\includegraphics[width=0.45\textwidth]{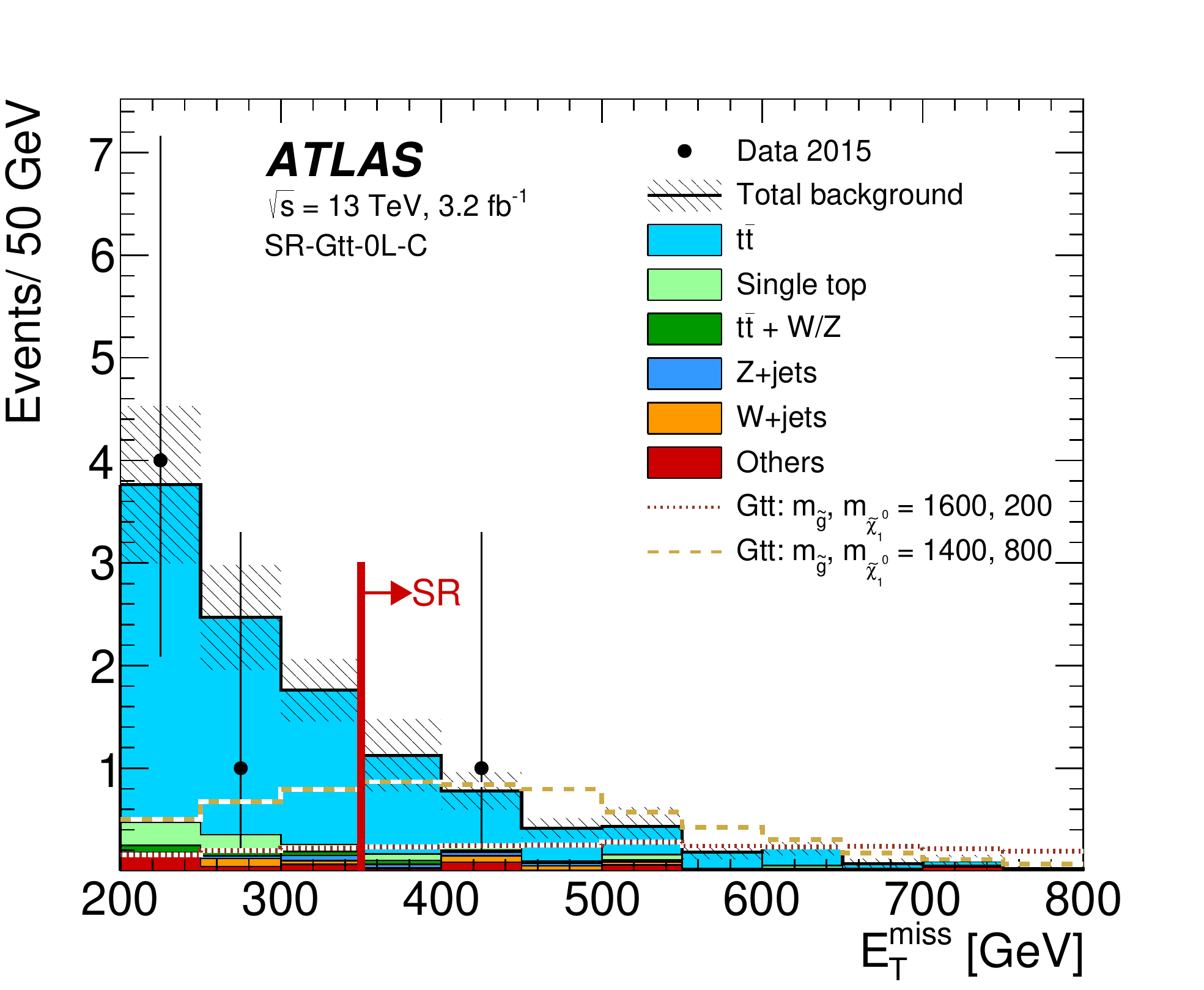}}
\subfigure[]{\includegraphics[width=0.45\textwidth]{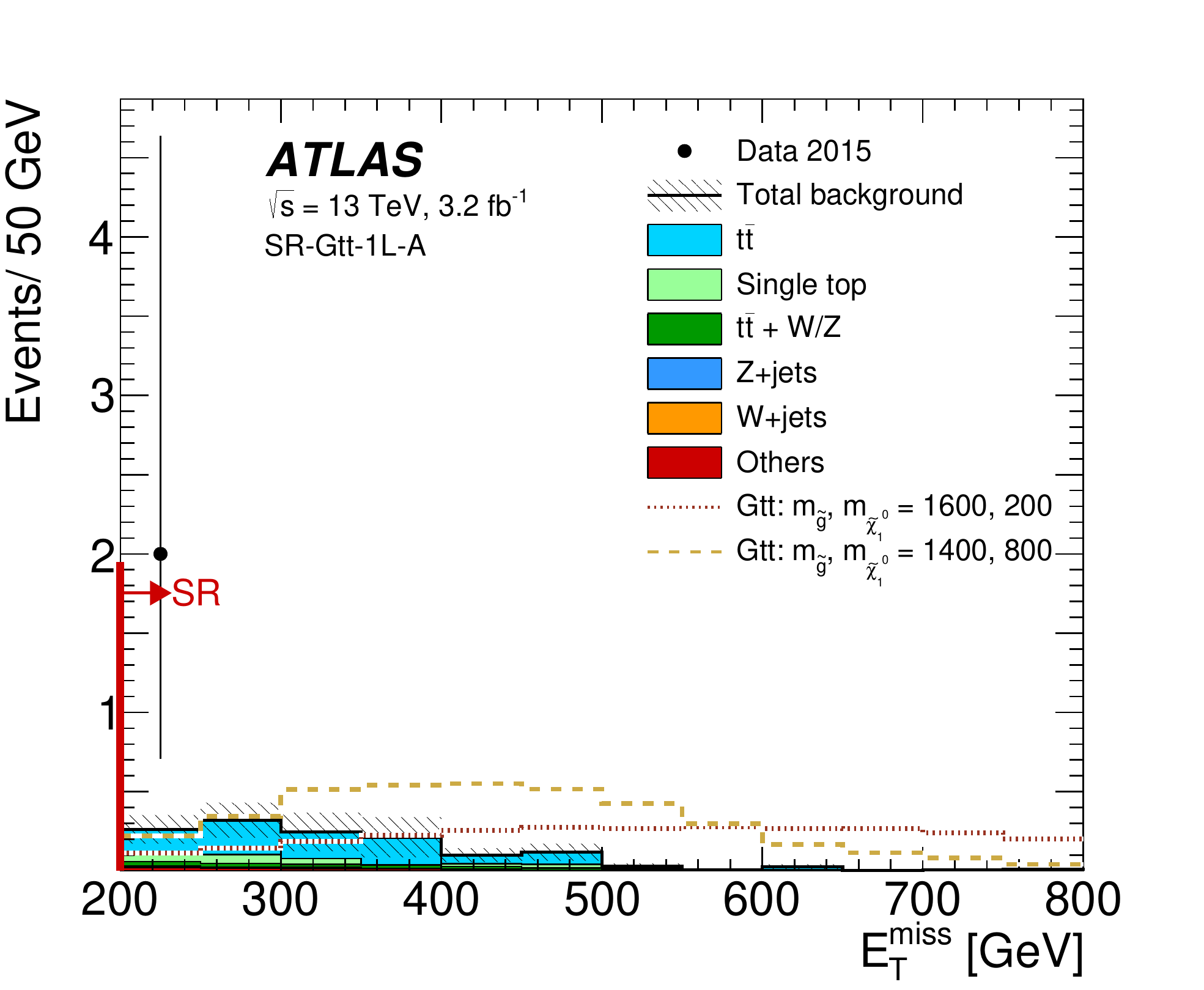}}
\caption{Distributions of $\met$ for
  (a) SR-Gbb-B, (b) SR-Gtt-0L-C and (c) SR-Gtt-1L-A. The $\met$ threshold is set to
200~GeV for these plots, with the red lines indicating the
threshold values in the actual signal regions for SR-Gbb-B and SR-Gtt-0L-C (the $\met$ threshold in
 SR-Gtt-1L-A is 200~GeV). The statistical and experimental
systematic uncertainties are included in the uncertainty band.   Two example signal models
          are overlaid. 
} \label{fig:SRplots}
\end{figure}

\section{Interpretation}
\label{sec:interpretation}

Since no significant excess over the expected background from SM processes is observed, the data 
are used to derive one-sided upper limits at 95\%~CL. Model-independent limits 
on the number of beyond-the-SM (BSM) events for each signal region are derived with pseudoexperiments using the CL$_s$ 
prescription \cite{Read:2002hq}.
They can 
be translated into upper limits on the visible BSM cross-section ($\sigma_\mathrm{vis}$), where $\sigma_\mathrm{vis}$ is
defined as the product of acceptance, reconstruction efficiency and production cross-section. The 
results are given in Table~\ref{mod-ind-lim}, where the observed ($S^{95}_\mathrm{obs}$) and expected
($S^{95}_\mathrm{exp}$) 95\% CL upper limits on the number of BSM
events are also provided.

\begin{table}
    \centering
{
\small
\begin{tabular*}{0.6\textwidth}{@{\extracolsep{\fill}}lccc}
\noalign{\smallskip}\toprule\noalign{\smallskip}
Signal channel                     & $\sigma_\mathrm{vis}$[fb]  &  $S_{\rm obs}^{95}$  & $S_{\rm exp}^{95}$   \\
\noalign{\smallskip}\midrule \noalign{\smallskip}
SR-Gbb-A & $0.94$ & $3.0$ & ${ 3.9 }^{ +1.3 }_{ -0.7 }$ \\ 
SR-Gbb-B & $0.94$ & $3.0$ & $ { 3.8 }^{ +1.4 }_{ -0.8 }$ \\ 
SR-Gbb-C & $1.74$ & $5.6$ & $ { 7.2 }^{ +2.6 }_{ -1.8 }$ \\

\noalign{\smallskip}\midrule\noalign{\smallskip}

SR-Gtt-1L-A & $1.49$ & $4.8$ & $ { 3.9 }^{ +1.4 }_{ -0.5 }$ \\ 
SR-Gtt-1L-B & $0.91$ & $3.0$ & $ { 3.0 }^{ +1.4 }_{ -0.0 }$ \\ 

\noalign{\smallskip}\midrule\noalign{\smallskip}

SR-Gtt-0L-A & $1.13$ & $3.6$ & $ { 4.4 }^{ +1.7 }_{ -1.0 }$ \\ 
SR-Gtt-0L-B & $1.16$ & $3.7$ & $ { 4.4 }^{ +1.9 }_{ -0.9 }$ \\ 
SR-Gtt-0L-C & $1.10$ & $3.5$ & $ { 4.5 }^{ +2.0 }_{ -1.2 }$ \\

\noalign{\smallskip}\bottomrule\noalign{\smallskip}
\end{tabular*}
}
\caption{The 95$\%$ CL upper limits on the visible cross-section
  ($\sigma_\mathrm{vis}$), defined as the product of acceptance, 
  reconstruction efficiency and production cross-section, and the observed and
  expected 95$\%$ CL upper limits on the number of BSM events ($S_{\rm
  obs}^{95}$ and $S_{\rm exp}^{95}$). }
   \label{mod-ind-lim}

\end{table}

The measurement is used to place exclusion limits on gluino and LSP masses in the Gbb and Gtt
simplified models. The results are obtained using the CL$_s$ prescription 
in the asymptotic approximation \cite{profilelh}.
The signal contamination 
in the control regions and the experimental systematic uncertainties
in the signal are taken into account for this calculation. 
For the Gbb models, the results are obtained from the Gbb signal region
with the best expected sensitivity at each point of the parameter
space of each model. For the Gtt models, the 0- and 1-lepton channels
both contribute to the sensitivity, and they are combined in a
simultaneous fit to enhance the sensitivity of the analysis. This is
performed by considering all possible permutations between the three
Gtt 0-lepton and the two Gtt 1-lepton signal regions for each point of
the parameter space, and the best expected combination is used. 
The 95\%~CL observed and expected exclusion limits for the Gbb and Gtt models are shown in the LSP and gluino mass plane in
Figures~\ref{fig:limits}(a) and (b), respectively. 
The $\pm1 \sigma^{\rm SUSY}_{\rm theory}$ lines around the observed limits are
obtained by changing the SUSY cross-section by one standard deviation
($\pm1\sigma$), as described in Section~\ref{sec:samples}.  The yellow band around the
expected limit shows the $\pm1\sigma$ uncertainty,  
including all statistical and systematic uncertainties 
except the theoretical uncertainties in the SUSY cross-section. It has been checked that 
the observed exclusion limits obtained from pseudoexperiments differ by less than 25 GeV 
from the asymptotic approximation in gluino or LSP mass in the combined limits in 
Figure~\ref{fig:limits}, although the difference can be up to 50 GeV when using single 
analysis regions. The two methods of computation produce equivalent expected limits. 

For the Gbb models, gluinos with masses below 1.78 \tev\
are excluded at 95\%~CL for LSP masses below 800~\gev. 
At high gluino masses, the exclusion limits are driven by the SR-Gbb-A
and SR-Gbb-B signal regions. 
The best
exclusion limit on the LSP mass is approximately 1.0 \tev, which
is reached for a gluino mass of
approximately 1.6 \tev. 
The exclusion limit is dominated by SR-Gbb-C for high LSP masses.
For the Gtt models,
gluino masses up to 1.8 \tev\ are excluded for massless LSP. For LSP
masses below 700~GeV, gluino masses below 1.76 \tev\ are
excluded.
For large gluino masses, the exclusion limits are driven by the
combination of SR-Gtt-1L-B and SR-Gtt-0L-A. 
 The LSP exclusion extends up to approximately
975~GeV, corresponding to a gluino mass of approximately $1.5
\tev$--$1.6 \tev$. 
The best exclusion limits are obtained by the combination of
SR-Gtt-1L-B and SR-Gtt-0L-C for high LSP masses. 
The ATLAS exclusion limits obtained with the full $\sqrt{s} = 8 \tev$
dataset are also shown in Figure~\ref{fig:limits}.
The current results largely improve on the $\sqrt{s} = 8 \tev$ limits despite the lower
integrated luminosity. The exclusion limit on the gluino
mass is extended by approximately $ 500 \gev$ and $400 \gev$ for the
Gbb and Gtt models for massless LSP, respectively. This improvement is primarily attributable to the
increased center-of-mass energy of the LHC. The addition of the IBL
pixel layer in Run~2, which improves the capability to tag
$b$-jets \cite{Capeans:1291633}, also particularly benefits this analysis that employs a
dataset requiring at least three $b$-tagged jets. The sensitivity of
the data analysis is also improved with respect to the $\sqrt{s}
= 8 \tev$ analysis \cite{3bjetsPaper} by using top-tagged large-$R$ jets, lepton isolation
adapted to a busy environment, and the $\mtb$ variable.

\begin{figure}[htbp]
\centering
\subfigure[]{\includegraphics[width=0.49\textwidth]{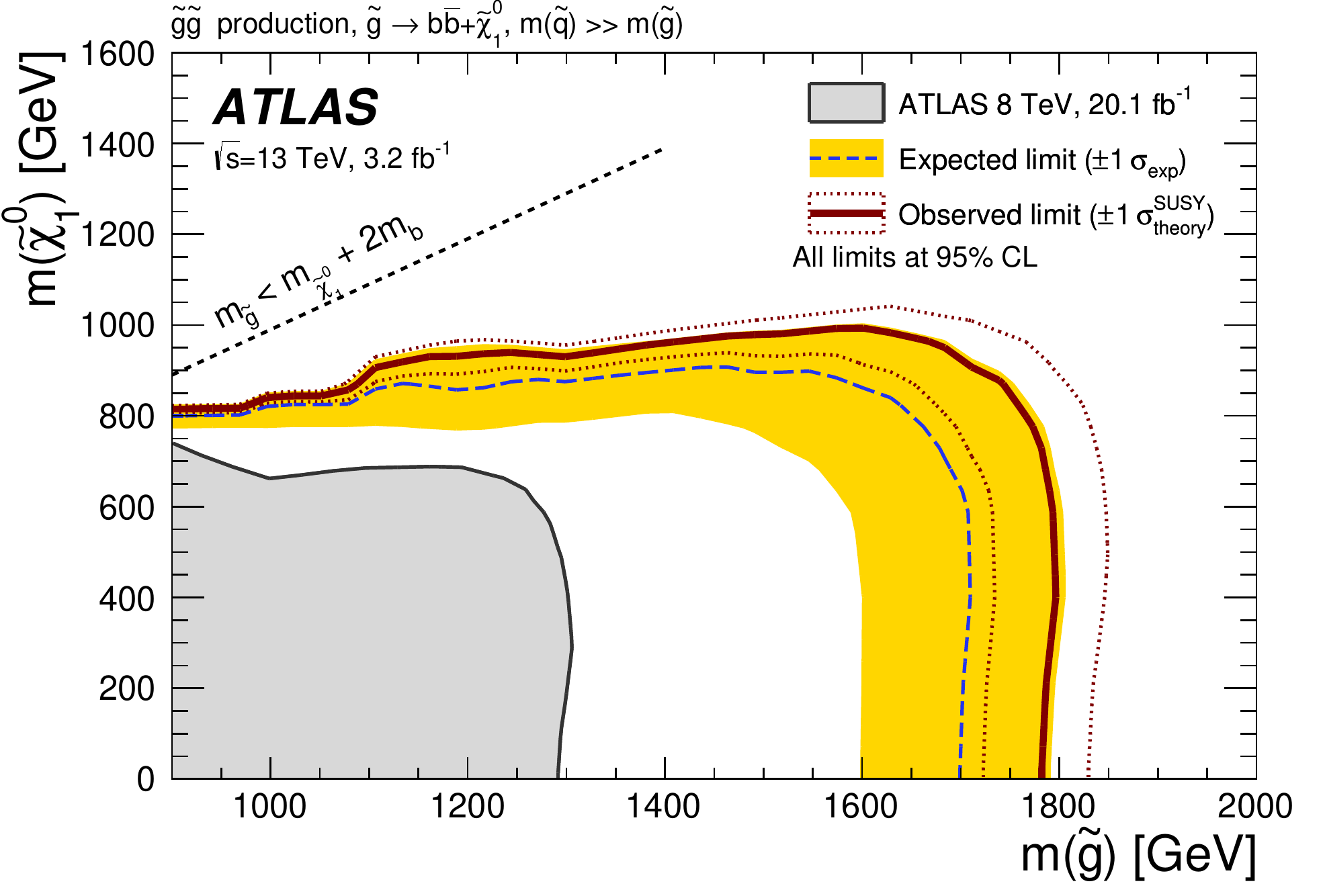}}
\subfigure[]{\includegraphics[width=0.49\textwidth]{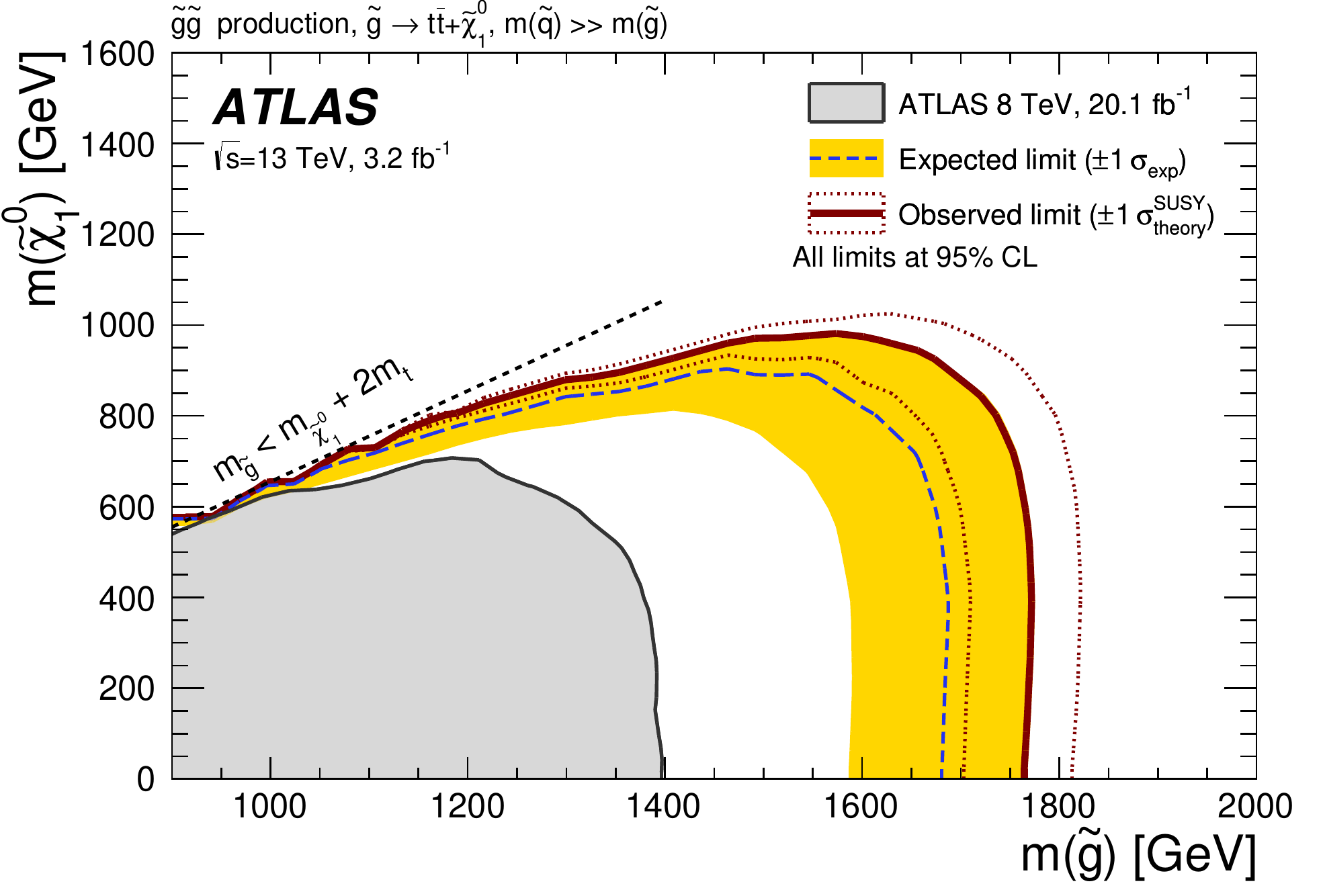}}
\caption{Exclusion limits in the $\ninoone$ and $\gluino$ mass plane
  for the (a) Gbb and (b) Gtt models. The dashed and solid bold lines
  show the 95\% CL expected and observed limits, respectively. The
  shaded bands around the expected limits show the impact of the
  experimental and background theoretical uncertainties. The dotted
  lines show the impact on the observed limit of the variation of the
  nominal signal cross-section by $\pm 1 \sigma$ of its theoretical
  uncertainty. The 95\%~CL observed limits from the
  $\sqrt{s} = 8 \tev$ ATLAS search requiring at least three $b$-tagged jets
  \cite{3bjetsPaper} are also shown. }
 \label{fig:limits}
\end{figure}

\section{Conclusion}
\label{sec:conclusion}
A search for pair-produced gluinos decaying via sbottom or stop is presented. LHC proton--proton 
collision data from the full 2015 data-taking period were analyzed, corresponding to an integrated 
luminosity of 3.2~\ifb\ collected at $\rts =13 \tev$ by the ATLAS detector. Several signal regions 
are designed for different scenarios of gluino and LSP masses. They require several high-$\pt$ jets, 
of which at least three must be $b$-tagged, large $\met$ and either zero or at least one charged lepton. 
For the gluino models with stop-mediated decays in which there is a large mass difference between 
the gluino and the LSP, large-$R$ jets identified as originating from highly boosted top quarks are employed.
The background is dominated by $\ttbar$+jets, which is normalized in dedicated control regions. 
No excess is found above the predicted background of each signal region. Model-independent limits are 
set on the visible cross-section for new physics processes. Exclusion limits are set on gluino and LSP 
masses in the simplified gluino models with stop-mediated and sbottom-mediated decays. For LSP masses 
below approximately 700~\gev, gluino masses of less than 1.78~\tev\ and 1.76~\tev\ are excluded at the 
95\% CL for the gluino models with sbottom-mediated and stop-mediated decays, respectively. These results 
significantly extend the exclusion limits obtained with the $\rts = 8 \tev$ dataset.

\section*{Acknowledgments}

We thank CERN for the very successful operation of the LHC, as well as the
support staff from our institutions without whom ATLAS could not be
operated efficiently.

We acknowledge the support of ANPCyT, Argentina; YerPhI, Armenia; ARC, Australia; BMWFW and FWF, Austria; ANAS, Azerbaijan; SSTC, Belarus; CNPq and FAPESP, Brazil; NSERC, NRC and CFI, 
Canada; CERN; CONICYT, Chile; CAS, MOST and NSFC, China; COLCIENCIAS, Colombia; MSMT CR, MPO CR and VSC CR, Czech Republic; DNRF and DNSRC, Denmark; IN2P3-CNRS, CEA-DSM/IRFU, France; 
GNSF, Georgia; BMBF, HGF, and MPG, Germany; GSRT, Greece; RGC, Hong Kong SAR, China; ISF, I-CORE and Benoziyo Center, Israel; INFN, Italy; MEXT and JSPS, Japan; CNRST, Morocco; FOM and 
NWO, Netherlands; RCN, Norway; MNiSW and NCN, Poland; FCT, Portugal; MNE/IFA, Romania; MES of Russia and NRC KI, Russian Federation; JINR; MESTD, Serbia; MSSR, Slovakia; ARRS and 
MIZ\v{S}, Slovenia; DST/NRF, South Africa; MINECO, Spain; SRC and Wallenberg Foundation, Sweden; SERI, SNSF and Cantons of Bern and Geneva, Switzerland; MOST, Taiwan; TAEK, Turkey; 
STFC, United Kingdom; DOE and NSF, United States of America. In addition, individual groups and members have received support from BCKDF, the Canada Council, CANARIE, CRC, Compute 
Canada, FQRNT, and the Ontario Innovation Trust, Canada; EPLANET, ERC, FP7, Horizon 2020 and Marie Sk{\l}odowska-Curie Actions, European Union; Investissements d'Avenir Labex and Idex, 
ANR, R{\'e}gion Auvergne and Fondation Partager le Savoir, France; DFG and AvH Foundation, Germany; Herakleitos, Thales and Aristeia programmes co-financed by EU-ESF and the Greek 
NSRF; BSF, GIF and Minerva, Israel; BRF, Norway; Generalitat de Catalunya, Generalitat Valenciana, Spain; the Royal Society and Leverhulme Trust, United Kingdom.

The crucial computing support from all WLCG partners is acknowledged
gratefully, in particular from CERN and the ATLAS Tier-1 facilities at
TRIUMF (Canada), NDGF (Denmark, Norway, Sweden), CC-IN2P3 (France),
KIT/GridKA (Germany), INFN-CNAF (Italy), NL-T1 (Netherlands), PIC (Spain),
ASGC (Taiwan), RAL (UK) and BNL (USA) and in the Tier-2 facilities
worldwide.

\printbibliography

\newpage
\input{atlas_authlist}

\end{document}

%% file: atlas_authlist.tex
\begin{flushleft}
{\Large The ATLAS Collaboration}

\bigskip

G.~Aad$^\textrm{\scriptsize 87}$,
B.~Abbott$^\textrm{\scriptsize 114}$,
J.~Abdallah$^\textrm{\scriptsize 65}$,
O.~Abdinov$^\textrm{\scriptsize 12}$,
B.~Abeloos$^\textrm{\scriptsize 118}$,
R.~Aben$^\textrm{\scriptsize 108}$,
O.S.~AbouZeid$^\textrm{\scriptsize 138}$,
N.L.~Abraham$^\textrm{\scriptsize 150}$,
H.~Abramowicz$^\textrm{\scriptsize 154}$,
H.~Abreu$^\textrm{\scriptsize 153}$
% ...
\bigskip
\\
% ...
$^{12}$ Institute of Physics, Azerbaijan Academy of Sciences, Baku, Azerbaijan\\
% ...
$^{65}$ University of Iowa, Iowa City IA, United States of America\\
% ...
$^{87}$ CPPM, Aix-Marseille Universit{\'e} and CNRS/IN2P3, Marseille, France\\
% ...
$^{108}$ Nikhef National Institute for Subatomic Physics and University of Amsterdam, Amsterdam, Netherlands\\
% ...
$^{114}$ Homer L. Dodge Department of Physics and Astronomy, University of Oklahoma, Norman OK, United States of America\\
% ...
$^{118}$ LAL, Univ. Paris-Sud, CNRS/IN2P3, Universit{\'e} Paris-Saclay, Orsay, France\\
% ...
$^{138}$ Santa Cruz Institute for Particle Physics, University of California Santa Cruz, Santa Cruz CA, United States of America\\
% ...
$^{150}$ Department of Physics and Astronomy, University of Sussex, Brighton, United Kingdom\\
% ...
$^{153}$ Department of Physics, Technion: Israel Institute of Technology, Haifa, Israel\\
$^{154}$ Raymond and Beverly Sackler School of Physics and Astronomy, Tel Aviv University, Tel Aviv, Israel\\
% ...
\end{flushleft}
